\documentclass[useAMS,usenatbib]{mnras}
\usepackage[english]{babel}
\usepackage{amsmath,amssymb}

\usepackage{graphicx}

\usepackage[normalem]{ulem}

\usepackage{verbatim}
\usepackage{color}
\usepackage{multirow}
\usepackage{mathtools}
\usepackage{epstopdf}

\usepackage{hyperref}
\usepackage{url}
\usepackage{xcolor}

\usepackage{gensymb}
\def\nat{Nature}

\def\prd{Phys. Rev. D}
\def\mnras{MNRAS}
\def\apj{ApJ}
\def\apjl{ApJL}
\def\apjs{ApJS}
\def\aap{A\&A}

\def\araa{ARA\&A}

\def\jcap{JCAP}

\def\be{\begin{equation}} 
\def\ee{\end{equation}} 
\def\ba{\begin{eqnarray}} 
\def\ea{\end{eqnarray}}

\def\msun{{\Msun}}

\def\HI{\hbox{H$\scriptstyle\rm I\ $}}

\def\gsim{\lower.5ex\hbox{\gtsima}} 
\def\lsim{\lower.5ex\hbox{\ltsima}} \def\gtsima{$\; \buildrel > \over 
\sim \;$} \def\ltsima{$\; \buildrel < \over \sim \;$} \def\prosima{$\; 
\buildrel \propto \over \sim \;$} \def\gsim{\lower.5ex\hbox{\gtsima}} 
\def\lsim{\lower.5ex\hbox{\ltsima}} 
\def\simgt{\lower.5ex\hbox{\gtsima}} 
\def\simlt{\lower.5ex\hbox{\ltsima}} 
\def\simpr{\lower.5ex\hbox{\prosima}} \def\la{\lsim} \def\ga{\gsim} 
  
 \def\gtsima{$\; \buildrel > \over \sim \;$} 
\def\ltsima{$\; \buildrel < \over \sim \;$} 
\def\gsim{\lower.5ex\hbox{\gtsima}} 
\def\lsim{\lower.5ex\hbox{\ltsima}} 
\def\simgt{\lower.5ex\hbox{\gtsima}} 
\def\simlt{\lower.5ex\hbox{\ltsima}} 
\def\simpr{\lower.5ex\hbox{\prosima}} 
\def\la{\lsim} 
\def\ga{\gsim}

\def\msun{\,{\rm \Msun}}

\def\E3{{\cal E}_{\rm g}^{III}}

\def\r12{r_{1/2}} 
\def\x12{x_{1/2}} 
\def\v12{v_{1/2}}

\def\HI{\hbox{H~$\scriptstyle\rm I $~}}

\def\nh2{n_{\rm H2}}
\def\fh2{f_{\rm H2}}

\def\angstrom{\textrm{A\kern -1.3ex\raisebox{0.6ex}{$^\circ$}}}

\def\msun{{\rm M}_{\odot}}

\makeatletter
\def\@hex@@Hex#1%
 {\if a#1A\else \if b#1B\else \if c#1C\else \if d#1D\else
  \if e#1E\else \if f#1F\else #1\fi\fi\fi\fi\fi\fi \@hex@Hex}
\makeatother
\definecolor{apcolor}{HTML}{b3003b}
\definecolor{cbcolor}{HTML}{ff0f00}
\definecolor{afcolor}{HTML}{b3443c}
\definecolor{ddcolor}{HTML}{077a2f}
\definecolor{fdcolor}{HTML}{0e3466}
\definecolor{sgcolor}{HTML}{008000}

\usepackage{etoolbox}
\makeatletter
\newcount\c@additionalboxlevel
\newcount\c@maxboxlevel
\patchcmd\@combinedblfloats{\box\@outputbox}{%
  \stepcounter{additionalboxlevel}%
  \box\@outputbox
  }

\AtBeginShipout{%
  \ifnum\value{additionalboxlevel}>\value{maxboxlevel}%
    \typeout{Warning: maxboxlevel might be too small, increase to %
      \the\value{additionalboxlevel}%
    }%
  \fi 
  \@whilenum\value{additionalboxlevel}<\value{maxboxlevel}\do{%
    \typeout{* Additional boxing of page `\thepage'}%
    \setbox\AtBeginShipoutBox=\hbox{\copy\AtBeginShipoutBox}%
    \stepcounter{additionalboxlevel}%
  }%
  \setcounter{additionalboxlevel}{0}%
}
\makeatother

\date{}
\pagerange{\pageref{firstpage}--\pageref{lastpage}} \pubyear{2022}

\title[Cosmic backgrounds from PBHs]{Cosmic radiation backgrounds from primordial black holes}

\author[F. Ziparo et al.]{F. Ziparo$^{1}$, S. Gallerani$^{1}$, A. Ferrara$^{1}$, F. Vito$^{2}$
\\
$^{1}$Scuola Normale Superiore, Piazza dei Cavalieri 7, I-56126 Pisa, Italy\\
$^{2}$INAF -- Osservatiorio di Astrofisica e Scienza dello Spazio di Bologna, Via Gobetti 93/3, I-40129 Bologna, Italy
}

\begin{document}

\maketitle
\label{firstpage}
\begin{abstract}
Recent measurements of the cosmic X-ray and radio backgrounds (CXB/CRB, respectively) obtained with Chandra and ARCADE2 report signals in excess of those expected from known sources, suggesting the presence of a yet undiscovered population of emitters. We investigate the hypothesis that such excesses are due to primordial black holes (PBHs) which may constitute a substantial fraction of dark matter (DM). We present a novel semi-analytical model which predicts X-ray and radio emission due to gas accretion onto PBHs, assuming that they are distributed both inside DM halos and in the intergalactic medium (IGM). Our model includes a self-consistent treatment of heating/ionization feedback on the surrounding environment. We find that (i) the emission from PBHs accreting in the IGM is subdominant at all times ($1\% \leq I_{\rm IGM}/I_{\rm tot} \leq 40\% $); (ii) most of the CXB/CRB emission comes from PBHs in DM mini-halos ($M_h \leq 10^6\ M_{\odot}$) at early epochs ($z>6$). While a small fraction ($f_{\rm PBH} \simeq 0.3\%$) of DM in the form of PBHs can account for the total observed CXB excess, the CRB one cannot be explained by PBHs. Our results set the strongest existing constraint on $ f_{\rm PBH} \leq 3\times 10^{-4}\ (30/M_{\rm PBH})$ in the mass range $1-1000\, M_\odot$. Finally, we comment on the implications of our results on the global \HI 21cm signal.
\end{abstract}


\maketitle

\section{Introduction}
The discovery of gravitational waves from black hole mergers detected with the LIGO/VIRGO interferometers provided us with a new channel to study astrophysical black holes. Interestingly, the generally high mass/low spin values inferred by these experiments \citep{Abbott16} are consistent with primordial black holes (PBHs) properties predicted by theoretical models (e.g. \citealt{sasaki16, Bird16, Clesse17}).

Primordial black holes (PBHs) are expected to be formed by gravitational collapse of overdense regions in the early Universe \citep{hawking74}, during the radiation dominated era. Given their early origin, PBHs can affect several properties of the Universe, such as the amplitude of primordial inhomogeneities \citep{Bellido17}, and the matter distribution in the early Universe (e.g., \citealt{Afshordi, Inman19}). In addition, extra energy injection due to PBHs accretion or evaporation affects the thermal history of the Universe \citep{Mena19, Junsong21}.

PBHs are non-baryonic Dark Matter candidates \citep{carr16}, and their abundance is described with the parameter $f_{\rm PBH}=\Omega_{\rm PBH}/\Omega_{\rm DM}$, i.e., the fraction of dark matter composed by PBHs. Under the standard assumption of a monochromatic PBH mass distribution, several constraints already exist on $f_{\rm PBH}$ from different astrophysical processes. The low-mass end of the PBH mass spectrum ($M<10^{-16}M_{\odot}$) is constrained through PBHs evaporation using the out-coming radiation as an observable. For instance, an upper limit of $10^{-8}<f_{\rm PBH}<1$ was found in the mass range $10^{-19}\msun< M_{\rm PBH}<10^{-16} \msun$ by comparing the gamma-ray radiation expected from PBH evaporation with the observed gamma-ray background \citep{Chen22}. Different constraints on $f_{\rm PBH}$ were set by the comparison between the BH merger rates detected by LIGO and the ones predicted by PBH models \citep{Kamionkowski17, Ballesteros18}. In the subsolar mass range ($10^{-8}\msun \leq M_{\rm PBH}\leq 1~\msun$), gravitational lensing has been the dominant process used to search for PBHs signature (e.g., \citealt{Tisserand07, Niikura19, Smyth20}) allowing for only a small fraction ($f_{\rm PBH} \simlt 10^{-2}$) of DM in the form of PBHs. 
The comparison between the CMB anisotropies value predicted when the contribution of accreting PBHs is accounted for and the observed one was used by \citet{Poulin17} and \citet{Ali-haimoud17} to set constraints ($f_{\rm PBH} < 10^{-3}$) in the mass range $10^2\leq M_{\rm PBH}\ \leq 10^5$. For higher masses, the dynamical effects induced by PBHs (e.g. tidal disruption of dwarf galaxies) dominate the constraints (e.g., \citealt{Monroy14, CarrSilk18, Zoutendijk20} allowing only a small fraction of DM ($f_{\rm PBH}\leq 10^{-5}$) to be comprised of PBHs.  

The entire mass spectrum is thus widely constrained, leaving only one window (i.e., $10^{-16}-10^{-11} M_{\odot}$, \citealt{carr20}, corresponding to the asteroid mass range) in which PBHs could constitute the entirety of dark matter. Although adopting a broader mass distribution allows the entirety of DM to be comprised of PBHs without exceeding existing constraints \citep{Hasinger2020}, such an approach requires a physically unjustified fine-tuning of the mass distribution. 

The presence of a (still unrecovered) population of accreting black holes at high redshift can be used to explain several observational results. For instance, a signal in excess of that produced by known sources was measured by several works \citep[e.g.][]{Hickox06, Hickox07, Cappelluti17} in the Cosmic X-ray Background (CXB) using the Chandra space telescope. The origin of such excess has been ascribed to the presence of still undiscovered, high-$z$ BHs \citep[$z>6$,][]{Salvaterra12, Kashlinsky16, Ananna20}.

The Absolute Radiometer for Cosmology, Astrophysics, and Diffuse Emission (ARCADE2) experiment measured measured an absolute sky brightness of $T_b= 54 \pm 6\ \rm mK$ at $3.3\rm~ GHz$. This implies an excess in the Cosmic Radio Background (CRB) of $5\times$ with respect to prediction from theoretical models \citep{ARCADE2}. In particular, the ARCADE2 results suggested the possibility that the CMB may not be the dominant source of the extragalactic radio background \citep{Ewall20}. The measurement was used, together with previous ones, to fit the radio-excess with a power-law spectrum, finding a spectral dependence $\nu^{-0.6}$ typical of the synchrotron radiation. The power-law results in 480 mK at 1.4 GHz or 21 cm.

PBHs can also explain the anomalous 21-cm absorption signal reported by the Experiment to Detect the Global EoR Signature \citep[EDGES, ][]{Bowman18}. 
EDGES aims at measuring single dipole signal in the sky-averaged global signal of the $21\rm cm$ radiation, in order to trace the volume-averaged ionization and thermal IGM evolution with redshift. This experiment has reported the detection of a strong absorption signal with a depth that is almost twice as strong as the expectation from the contribution of the Cosmic Microwave Background (CMB) alone \citep{Bowman18}. Given the expression for the differential brightness temperature $\delta T_b$ \citep{ewall}:
\begin{align}
\delta T_b \propto \left(1-\frac{T_{\rm CMB}+T_\gamma}{T_s}\right),
\end{align}
where $T_{\rm CMB}$ is the CMB brightness temperature, $T_s$ is the spin temperature, and $T_{\gamma}$ is a the brightness temperature of an extra component in the radio background. Two theoretical explanation have been proposed for the \citet{Bowman18} result: the first one is that the $T_s$ value is lower than expected at $z \sim 17$. This hypothesis has been explored by different works \citep{barkana18, loeb18, fialkov18}; the second one is the presence of an additional contribution to the CRB as suggested by \citet{EDGES} and other works e.g. \citet{ewall, radioloud21}. 

In this work, we explore the possiblity that PBHs are the sources of the excess signals in the CXB and CBR, and can explain the EDGES absorption signal. A widely used approach to model PBH accretion in the eary Universe (e.g., \citealt{ricotti08, ricotti09, Mena19, Hasinger2020}) is to assume that PBHs are surrounded by a uniform gas distribution, thus neglecting accretion from denser gas, located in virialized structures. In this work we add this important ingredient, i.e. PBH accretion inside DM halos. A similar approach was undertaken by \citet{gaggero17} in a Milky Way context. We build up on that work, further modeling the PBH-gas interaction in a cosmological framework. 

The paper is organized as follows. In Sec. \ref{model} we describe the model\footnote{We adopt a $\Lambda$CDM cosmology in agreement with Planck18 \citep{Planck18} results: $\Omega_{\rm m}=0.315$, 
$\Omega_{\rm \Lambda}=0.685$, $\Omega_{\rm b}=0.049$, $\sigma_8 = 0.811$, $n_s=0.965$, 
and $H_{\rm 0}=100 \,h$~km~s$^{-1}$~Mpc$^{-1}=67.4$~km~s$^{-1}$~Mpc$^{-1}$.} adopted in this work, in Sec. \ref{heation} we review the model implications for the IGM thermal history; Sec. \ref{results} contains the main findings of this work. In Sec. \ref{PBHconst} we compare our results with previous ones, and Sec. \ref{conclusion} presents a brief summary.

\section{Model}\label{model}
In this section, we describe the theoretical model adopted to distribute PBHs in the Universe (Sec. 2.1), to compute their accretion (Sec. 2.2), the resulting X-ray and radio emission (Sec. 2.3), and the corresponding cosmic backgrounds (Sec. 2.4). 

%Our aim is to model PBHs accretion inside halos and into the IGM and to describe how this process affect the evolution of the Universe. We first describe how PBHs populate the IGM and the boundary conditions of the accretion process in this environment. Then we describe how dark matter halos are modeled and how PBHs accreet in virialized object. Lastly we describe how the accretion process can result in X-ray and radio luminosity.
%\subsection{PBHs cosmological abundanceS?}
\subsection{PBHs cosmological distribution}
In this work, we assume that PBHs constitute a fraction of dark matter $f_{\rm PBH}=\Omega_{\rm PBH}/\Omega_{\rm DM}$. The PBH density distribution is assumed to track dark matter, which is in turn organised into virialized objects (halos) and a diffuse component at the mean cosmic density, i.e. the intergalactic medium (IGM). The PBH number density, $n_{\rm PBH}$, at redshift $z$ is then given by:
\begin{equation}
n_{\rm PBH}(z)=\frac{\Omega_{\rm DM}\rho_c f_{\rm PBH}(1+z)^3}{M_{\rm PBH}}=n_{\rm PBH}^{\rm IGM}(z)+n_{\rm PBH}^{h}(z),  
\end{equation}
where $\rho_c = 3 H^2/8 \pi G$ is the critical density of the Universe, $M_{\rm PBH}$ is the PBH mass, and $n_{\rm PBH}^{\rm IGM}$ ($n_{\rm PBH}^{h}$) represents the number density of PBHs in the IGM (DM halos). We assume a monochromatic PBH mass distribution, in the range $1 \leq M_{\rm PBH}\leq 1000 ~\msun$.

The relative abundance of PBHs in halos and IGM depends on the fraction of dark matter that collapses into virialized structures per comoving volume, that can be computed through the Press-Schechter formalism \citep{PS74}:
\begin{equation}
  f_{\rm coll}(\> M_{\rm min}, z)=\mathrm{erfc}\left[\frac{\delta_c(z)}{\sqrt{2}\sigma_M} \right],
\end{equation}
where $\delta_c=1.686/D(z)$ is the critical density for collapse, $D(z)\propto (1+z)^{-1}$ is the growth factor, $\sigma_M$ is the standard deviation of the linearly extrapolated matter power spectrum, and $M_{\rm min}$ is the minimum halo mass such that efficient cooling processes are triggered (see Sec. \ref{backint}).
%In this work we assume that a fraction $f_{\rm PBH}=\Omega_{\rm PBH}/\Omega_{\rm DM}$ of dark matter is composed of PBHs. We take into account for two populations of PBHs: one residing into the IGM, and the second contained into dark matter halos. The fraction of dark matter not embedded into virialized structures per comoving volume is a function of redshift, 
\subsubsection{PBHs in the IGM}
The number density of PBHs in the IGM is given by: 
\begin{equation}\label{n_h}
  n_{\rm PBH}^{\rm IGM}(z)=\frac{\Omega_{\rm DM}\rho_c(1+z)^3(1-f_{\rm coll}(z))f_{\rm PBH}}{M_{\rm PBH}}.
\end{equation}
This expression shows that $n_{\rm PBH}^{\rm IGM}$ decreases with decreasing redshift. Besides cosmic expansion, this redshift evolution is due to the on-going structure formation, as more dark matter falls onto virialized halos, more PBHs are locked into DM halos than in the IGM.

\subsubsection{PBHs in dark matter halos}
The total amount of PBHs inside dark matter halos can be expressed, as a function of redshift, as:
\begin{equation}\label{n_igm}
  n_{\rm PBH}^{h}(z)=\frac{\Omega_{\rm DM}\rho_c(1+z)^3f_{\rm coll}(z)f_{\rm PBH}}{M_{\rm PBH}}.
\end{equation}
To distribute PBHs into a DM halo of mass $M_{\rm vir}$, we assume that they follow a Navarro, Frenk \& White \citep[NFW,][]{NFW97} density profile:
\begin{equation}
\rho_{\rm DM}(r)=\frac{\rho_c\delta_c}{cx(1+cx)^2},
\end{equation}
where $x=r/r_{\rm vir}$ denotes the radial distance in units of the virial radius $r_{\rm vir}$ \citep{barkana2001beginning}:
\begin{equation}
\begin{split}
  r_{\rm vir}=\ &0.784\ \left(\frac{M_{\rm vir}}{10^8\ h^{-1}\ M_{\rm \odot}}\right)^{1/3}\left[\frac{\Omega_m}{\Omega^z_m}\frac{\Delta_c}{18\pi^2}\right]^{-1/3}\ \\ 
  &\left(\frac{1+z}{10}\right)^{-1}h^{-1}\ \mathrm{kpc},
\end{split}
\end{equation}
 where the overdensity relative to $\rho_c$ at the collapse redshift can be expressed as $\Delta_c=18\pi^2+82d-39d^2$, with $d=\Omega_m^z-1$ and $\Omega_m^z=\Omega_m(1+z)^3/(\Omega_m(1+z)^3+\Omega_{\Lambda})$; $\Delta_c$ is related to $\delta_c$ through the following relation:

\begin{equation}\label{overD}
\delta_c=\frac{\Delta_c}{3}\frac{c^3}{ln(1+c)-c/(1+c)}=\frac{\Delta_c}{3}\frac{c^3}{F(c)},
\end{equation}
where $c$ is the concentration parameter that depends on $M_{\rm vir}$:
\begin{equation}
\log c=1.071-0.098(\log M_{\rm vir}-12).
\label{maccio}
\end{equation}
The last relation is the result of N-body simulations developed by \citet{maccio07}; in Appendix \ref{concentration}, we discuss its applicability to low-mass ($M<10^8 \msun$)and high-redshift ($z>6$) objects.

\begin{figure}
\centering
\includegraphics[width=0.45\textwidth]{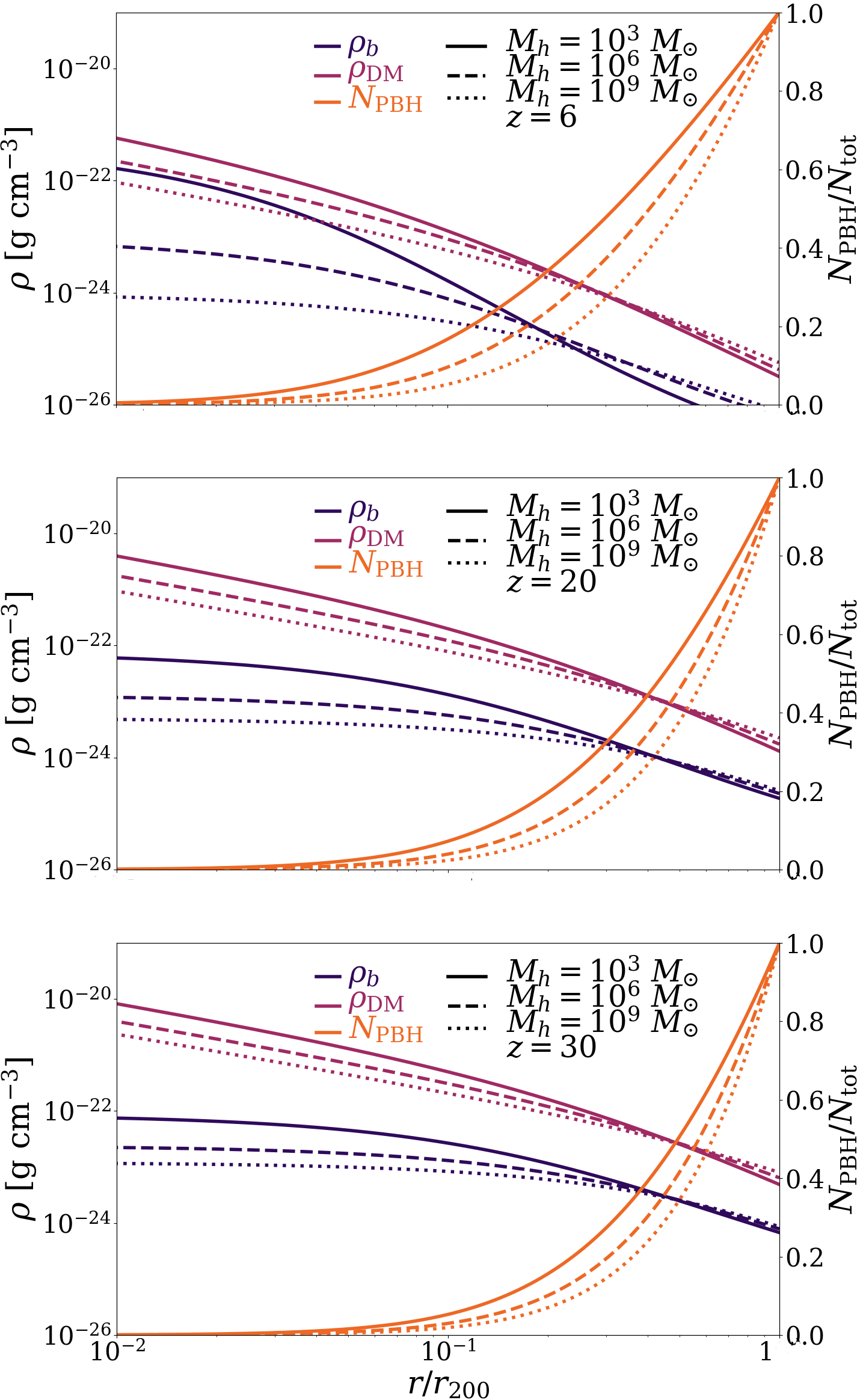}
\caption{Baryon density, dark matter density, and PBHs integrated number as a function of the virial radius. Lines correspond to different halo masses. The solid, dashed, dotted line stays for a halo of mass $M_h=10^3 \msun$, $M_h=10^{6} \msun$, $M_h=10^{9} \msun$, respectively. Different colors represent different quantities: orange PBHs cumulative number, violet baryon density, purple dark matter density. The 3 panels present 3 redshifts: $z=6, 20, 30$. PBHs mass is set at $30 M_{\odot}$.}\label{NUMDENS}
\end{figure}

The total number of PBHs inside the halo can be computed by imposing that the halo mass is constituted of a fraction $f_{\rm PBH}$ of PBHs:
\begin{equation}\label{ntot}
N_{\rm tot}=f_{\rm PBH}\left(\frac{M_h}{M_{\rm PBH}}\right).
\end{equation}
The number of PBHs within any given radial distance is then given by:
\begin{equation}
N_{\rm PBH}(r)=f_{\rm PBH}\frac{4\pi}{M_{\rm PBH}}\int^{r}_{\rm 0}\rho_{\rm DM}(r')r'^2dr'.
\end{equation}
%The number of black holes in the halo as a function of the radius follow the matter profile. The evolution of the PBHs number with the virial radius is shown in figure \ref{NUMDENS}. \zip{rileggi}We show the evolution with the virial radius of the baryon density (purple), and dark matter density (magenta) on the y1 axes. On y2 axes we show the cumulative number of PBHs. It is clear that from the figure that the vast majority of PBHs sits far from the center of the halo, according to the NFW mass distribution. Moving to lower redshift shift the middle point, $N_{PBH}= 1/2 N_{tot}$, towards smaller radii. This effect is attributed to the evolution of the concentration parameter with the redshift (see appendix \ref{concentration}).
Fig. \ref{NUMDENS} shows the baryonic and DM density distributions (left axis), and the number of PBHs normalized to their total number (right axis), as a function of the radial distance from the center of the halo. According to our formalism, at a given redshift, less massive halos ($M_{\rm vir}=10^3 M_{\odot}$) are more concentrated than more massive ones ($M_{\rm vir}=10^9 M_{\odot}$); thus, 50\% of $N_{\rm PBH}$ reside at smaller radial distances in the less massive halos ($0.3~r_{\rm vir}$ vs. $0.5~r_{\rm vir}$). Similarly, at a given fixed virial mass (e.g. $M_{\rm vir}=10^6 M_{\odot}$), low redshift ($z=6$) DM halos are more concentrated than high redshift ($z=20$) ones. As a result, 50\% of $N_{\rm PBH}$ reside at smaller radial distances in the less massive halos ($0.4~r_{\rm vir}$ vs. $0.6~r_{\rm vir}$).

\begin{figure}
\centering
\includegraphics[width=0.45\textwidth]{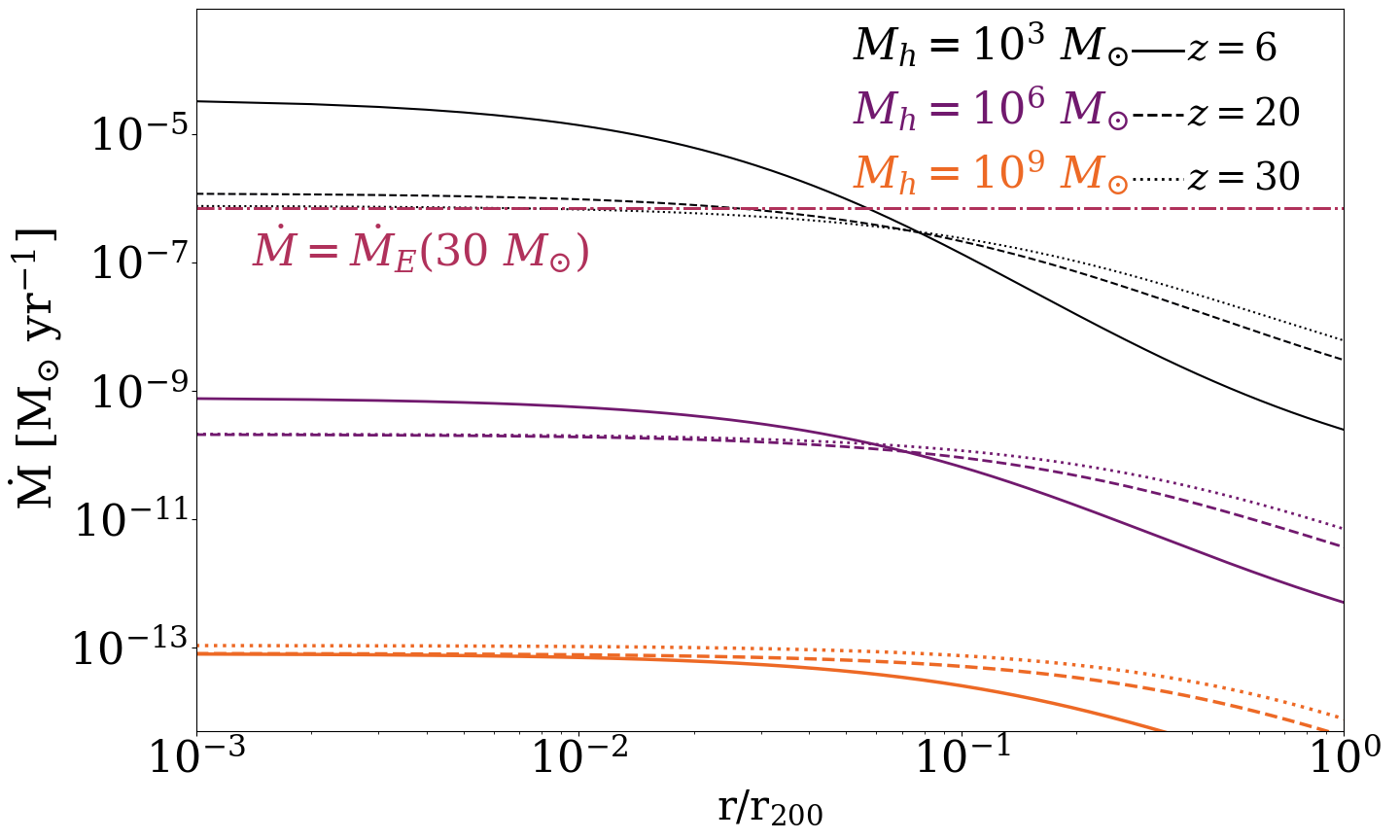}
\caption{Accretion rate (for a single $30\ M_{\odot}$ black hole) as a function of the virial radius. The solid (dashed/dotted) line represents the redshift $z=6$ ($z=20/z=30$). Different colors denote different halo masses: black $M_h=10^3 \msun$, violet $M_h=10^{6} \msun$, and orange $M_h=10^{9} \msun$. The horizontal purple line is the Eddington rate for a $30 M_{\odot}$ black hole.}\label{accretiionZ}
\end{figure}

\subsection{PBH accretion}
We assume that, both in the IGM and in halos, PBHs accrete according to the Bondi-Hoyle-Lyttelton model:
\begin{equation}\label{accrex}
\dot{M}_{\rm PBH}=\lambda \frac{4\pi G^2M_{\rm PBH}^2\rho_{\rm gas}}{(c^2_{\rm s}+v^2_{\rm BH})^{3/2}},
\end{equation}
where $G$ is the universal gravity constant, $\rho_{\rm gas}$ is the gas density, $c_s$ is the sound speed, $v_{\rm BH}$ is the relative velocity between the black hole and the gas, and $\lambda = 0.01$ is the accretion eigenvalue, that accounts for non-gravitational effects in the process \citep{Xie12}. We prevent the gas accretion to exceed the Eddington rate $\dot{M}_{E}=L_E/(\varepsilon~c^2)$, where $L_E$ is the Eddington luminosity:
\begin{equation}
L_{E}=\frac{4\pi GM m_p c}{\sigma_T}=1.3\times10^{38}\biggl(\frac{M_{\rm PBH}}{M_{\rm \odot}}\biggl)\ {\rm erg\ s^{-1}} ,
\end{equation}
and $\varepsilon=0.1$ is the assumed radiative efficiency\footnote{The value of the radiative efficiency is strongly dependent on the geometry of accretion, and can vary from $\varepsilon\ <\ 10^{-4}$, in the case of spherically symmetric accretion to $\varepsilon\ <\ 0.43$ for disk accretion processes \citep{Shakura73, shapiro73}}. 

PBH accretion in the IGM and inside halos proceeds in a considerably different manner \citep{ricotti08,deluca20}. In what follows we characterize $\rho_{\rm gas}$, $c_s$, and $v_{\rm BH}$ in these two cases, separately. \\

\subsubsection{PBH accretion in the IGM}
Given the number density in eq. \ref{n_igm}, the accretion from all the PBHs distributed in the IGM per comoving volume can be written as:
\begin{equation}
  \dot{M}(z)=4\pi \lambda G^2M_{\rm PBH}^2\frac{\rho_{\rm IGM}(z)}{[c_s(z)^2+v_{\rm rel}(z)^2]^{3/2}}n_{\rm PBH}^{\rm IGM}(z),
\end{equation}
where $\rho_{\rm IGM}$ is the gas density, $c_s$ is the sound speed in the IGM, and $v_{\rm rel}$ is the relative velocity between dark matter and baryons. 
Following \citet{ricotti09}, we assume that PBHs in the IGM are surrounded by a uniform distribution of gas with density:
\begin{equation}
  \rho_{\rm IGM}=250\, \mu m_p \biggl(\frac{1+z}{1000} \biggl)^3\ \mathrm{g\ cm^{-3}},
\end{equation}
where $m_p$ is the proton mass and $\mu = 1.22$ is the mean molecular weight of a neutral gas of primordial composition.
The IGM sound speed is 
\begin{equation}
c_s=( k_BT_{\rm IGM}/\mu m_p)^{1/2},
\end{equation}
where $k_B$ is the Boltzmann constant, and ${T_{\rm IGM}}$ is the gas temperature that depends on large scale effects, as discussed in Sec. \ref{heation}.

\begin{figure*}
\begin{center}
\includegraphics[width=1\textwidth]{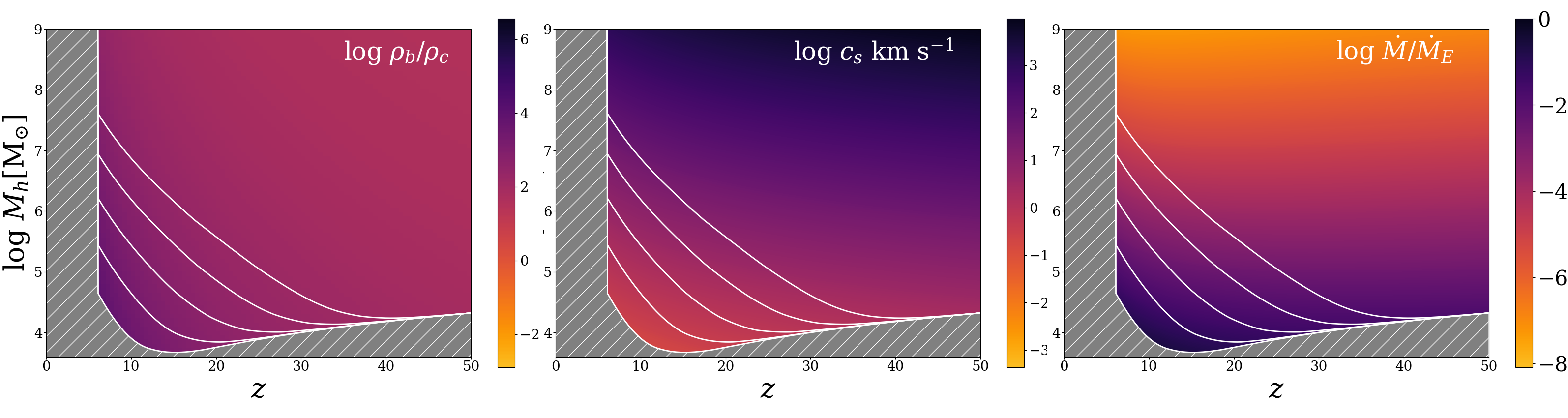}
\caption{{\it Left panel}: baryon density (normalized to the critical density, $\rho_c$) as a function of halo mass, $M_h$, and redshift, $z$. %as expected the baryon overdensities inside mini-halos stays constant around $\sim 100\ \rho_c$. 
{\it Middle}: same as left panel for sound speed, $c_s$. {\it Right}: same as left panel for the accretion rate of a $30\ M_{\odot}$ PBH (normalized to the Eddington rate). Note that accretion is more efficient in small halos at low redshift due to their higher density and lower sound speed. All values have been computed at $r=0.01\ r_{\rm vir}$, where the baryon density and accretion rate reach their maximum. In each panel, the hatched grey regions represents the excluded region of the parameter space. The continuous white lines are an extension of the excluded region for the different $M_{\rm min}$ considered. We discuss how to select $M_{\rm min}$ in Sec. \ref{heation}. %The continuous white lines follow the minimum mass evolution shown in Fig. \ref{IONTEMPEVO}, right panel. 
}\label{AccretionVariables}
\end{center}
\end{figure*}

For what concerns the relative velocity between dark matter and baryons, we adopt the relation proposed by \citet{Ali-haimoud17}:
\begin{equation}\label{rel_vel}
  v_{\rm rel}(z)\simeq 30 \min\biggl[1, \left(\frac{1+z}{1000}\right)\biggl]\, \mathrm{km\ s^{-1}}.
\end{equation}
\subsubsection{PBH accretion in halos}
Inside halos, the accretion rate strongly depends on the radial distance $r$:
\begin{equation}\label{accrex_halo}
\dot{M}_{\rm BHL}(r)=\lambda \frac{4\pi G^2M_{\rm PBH}^2\rho_{\rm gas}(r)}{(c^2_{\rm s}+v^2_{\rm BH})^{3/2}}.
\end{equation}
We adopt the gas density distribution from \cite{makino98}:
\begin{equation}\label{ro_bar}
\rho_{\rm gas}=\rho_0\ \exp\left\{-\frac{\mu m_p}{2k_BT_{\rm vir}}[v^2_e(0)-v^2_e(r)]\right\},
\end{equation}
where $\rho_0(M)$ is the central gas density\footnote{The constant $\rho_0$ is set such that $M_b=({\Omega_b}/{\Omega_m})M_h$.}, $T_{\rm vir}$ is the virial temperature of the halo:
\begin{equation}\label{Tvir}
\begin{split}
T_{\rm vir}=\ & 1.98\times 10^4\ \left(\frac{\mu}{0.6}\right)\ \left(\frac{M_{\rm vir}}{10^8\ h^{-1}\ M_{\rm \odot}}\right)^{2/3}\ \\ & \left[\frac{\Omega_m}{\Omega^z_m}\frac{\Delta_c}{18\pi^2}\right]^{1/3}\left(\frac{1+z}{10}\right)\ \rm K,
\end{split}
\end{equation}
and $v_e$ is the escape velocity that can be obtained through the following relations:
\begin{align*}
M(r)&=\int^{r}_{0}4\pi \rho(r')dr'=M_{\rm vir}\frac{F(cx)}{F(c)},\\
%\end{equation}
%\begin{equation}
v^2_c(r)&=\frac{GM(r)}{r}=v^2_{\rm c,vir}\frac{F(cx)}{F(c)},\\
%\end{equation}
%\begin{equation} 
v^2_e(r)&=2\int^{r_{vir}}_{r}\frac{GM(r')}{r'^2}dr'\approx 2v^2_c\frac{F(cx)+\frac{cx}{1+cx}}{xF(cx)},\\
\end{align*}
where $F(c)$ is defined in Eq. \ref{overD}, and $v^2_{\rm c,vir}=GM_{\rm vir}/r_{\rm vir}$ is the circular velocity. We further assume that inside the halo the sound speed is constant and equal to:
\begin{equation}\label{cs}
c_s=( k_BT_{\rm vir}/\mu m_p)^{1/2} {\rm km\ s^{-1}}=8.3\ T^{1/2}_{\rm vir,4}\ \rm km\ \rm {s}^{-1},
\end{equation}
where $T_{\rm vir,4}$ is the virial temperature in units of $10^4\ \rm K$. We finally set to zero the relative velocity between PBHs and baryons, assuming that dark matter and baryons are in hydrodynamical equilibrium.

In Fig. \ref{accretiionZ}, we show how the accretion rate varies within the virial radius, for different halo masses and redshifts. For a fixed halo mass, the accretion rate is higher at low redshift; for a fixed redshift, the accretion rate is higher for less massive halos. To understand these trends, in the left and middle panels of Fig. \ref{AccretionVariables} we plot the redshift and $M_h$ dependence of the gas density and sound speed, which govern the PBH accretion rate (eq. \ref{accrex_halo}). For illustration, both quantities are computed at $r=0.01 r_{\rm vir}$. The baryon overdensity $\rho_b/\rho_c$ inside halos remains almost constant around a value $\simeq 100$ along cosmic history, but slightly increases for low $z$ and/or small DM halos. Instead, the sound speed decreases both with halo mass and redshift. 

These two combined effects, explain the results shown in the right panel of Fig. \ref{accretiionZ}, where the accretion rate at $r=0.01\ r_{\rm vir}$ is shown as a function of $M_h$ and $z$. Accretion is more efficient in the central region of low mass halos ($M \leq 10^6 M_{\odot}$), at any given $z$. In these objects, accretion is so efficient that it compensates the reduced total number of PBHs that scales linearly with $M_h$ (Eq. \ref{ntot}). Given that small halos are also more numerous relative to more massive ones, we expect their PBH emission to be dominant.

\subsection{X-ray and radio luminosity}\label{luminosity}
Given the bolometric luminosity $L_{\rm B}=\varepsilon \dot{M} c^2$ of an accreting PBH, the X-ray luminosity ($L_X$) can be computed by means of a bolometric correction $f_X$, i.e.
\begin{equation}
L_X=f_XL_{\rm B}.
\end{equation}
Following \citet{ewall}, we assume $f_X=0.1$ in the $2-10\ \mathrm{keV}$ band \citep[see also][]{Poulin17, Mena19, 21cmForestPBH}. We model the X-ray spectra of accreting PBHs with a power-law with an exponential cut-off at high energies \citep{Poulin17, Mena19, 21cmForestPBH}, and compute the specific luminosity in the X-ray band ($L_{X,\nu}$) as:
\begin{equation}
  L_{X,\nu} \propto {\nu}^{-\alpha}e^{(-\nu/\nu_{\rm cut})},
\end{equation}
where\footnote{Typical values of spectral indices in the X-ray band are in the range $\alpha = [0.7, 1.3]$ \citep{Yuan14}) and $\nu_{\rm cut}$ is the cut-off frequency, such that $h\nu_{\rm cut}=200\ \mathrm{keV}$.} $\alpha=0.7$ is the spectral index in the X-ray band, and $\nu_{cut}$ is the cut-off frequency.

For what concerns the radio luminosity $L_R$, we assume the fundamental plane relation \citep{Gallo03, merloni03, wang06}. In particular, following \citet{Hasinger2020}, we adopt the relation found by \cite{wang06}, based on the radio luminosity measured at 1.4 GHz and the X-ray luminosity in the 0.1-2.4 keV band: 
%A proportionality between X-ray and radio luminosity ( $L_R \sim L_X^{0.7}$) was observed in X-ray binaries . The fundamental plane relation was also studied in \cite{merloni03}, who found a steeper slope that better reproduced the scaling of the radio luminosity with the black hole mass and a proportionality the 0.5-2 keV luminosity.
%The same relation was studied in , the 1.4 GHz radio luminosity radio $L_{\rm R}$ is obtained from the X-ray luminosity in the 0.1-2.4 keV band and the black hole mass. In this work we use the latter, following :
\begin{equation}
\begin{split}
  & \log\left(\frac{L_R}{10^{40}\mathrm{erg\ s^{-1}}}\right) = \\ & = 0.85\log\left(\frac{L_X}{10^{44}\mathrm{erg\ s^{-1}}}\right)+0.12\log\left(\frac{M_{\rm PBH}}{10^{8}\mathrm{M_{\rm \odot}}}\right).
\end{split}
\end{equation}

We finally assume for the radio luminosity a typical spectral index of syncrothron emission, $L_{R,\nu}\propto \nu^{-0.6}$.

\subsection{Background intensity}\label{backint}
The comoving specific emissivity due to PBHs accreting in the IGM can be written as:
%\begin{equation}
 %  \dot{\rho}_{\rm IGM}(\nu, z)= L_{\rm rad}(z)\biggl(\frac{\nu}{\nu_0}\biggl)^{-0.6}n_{\rm IGM}(z)\ \mathrm{erg\ s^{-1}\ Mpc^{-3} Hz^1},
%\end{equation}
%where $L_{\rm rad}(z)$ is the radio luminosity of a single PBH at a given redshift. 
\begin{equation}
  \dot{\rho}_{\rm IGM}(\nu, z)= L_{\nu}(z) n_{\rm IGM}(z)\ \mathrm{erg\ s^{-1}\ Mpc^{-3} Hz^{-1}}.
\end{equation}
where $\nu$ is the (rest-frame) frequency, and $L_{\nu}$ represents the X-ray or radio specific luminosity computed with the formalism described in the previous section.
For what concerns the emissivity from halos, $\dot{\rho}_h$, we first compute the integrated luminosity from a single object of mass $M_{\rm vir}$:
\begin{equation}\label{halum}
L_{\nu}(M_{\rm vir})= \int^{r_{\rm vir}}_{\rm 0}L_{\nu}(r')n_{\rm PBH}(r')dr',
\end{equation}
and then we integrate over the halo mass function \citep{HMFcalculator}:
\begin{equation}\label{bhint}
\dot{\rho}_{h}(\nu, z)=\int^{M_{\rm max}}_{\rm M_{\rm min}}L_{\nu}(M', z)\ n_h(M', z)dM',
\end{equation}
where $n_h(M, z)$ gives the number of halos of mass $M$ at redshift $z$, $M_{\rm max} = M_h(T_{\rm vir}=10^4 \mathrm{K})$ is the minimum DM halo mass for star formation to happen, and $M_{\rm min}$ is the minimum mass of DM halos that can virialise and form a baryon overdensity:
\begin{equation}
  M_{\rm min}(T_{\rm IGM}, z)=1.3\times 10^3 M_{\rm \odot} \left(\frac{10}{1+z}\right)^{3/2} \left(\frac{T_{\rm IGM}(z)}{1\ \mathrm{K}}\right)^{3/2}.
\end{equation}
The latter depends on the IGM temperature evolution and will be computed in Sec. \ref{neutral_regions}.
%The previous expression allow us to discriminate between mini-halos that contain gas, and the one, at lower masses that are only composed of dark matter. Halos at a temperature lower then the IGM temperature are unable to virialize gas, which is crucial for the accretion process.

Finally, we proceed to compute the background intensity, $I_{\rm PBH}$, as follows:
\begin{equation}\label{intensity}
I_{\rm PBH}(\nu, z)=\frac{c(1+z)^3}{4\pi \nu}\int_z^{\infty}\frac{\dot{\rho}(\nu, z')}{(1+z')H(z')}dz',
\end{equation} 
where $H(z)$ is the Hubble parameter. Usually, in the radio frequency range the background intensity, $I$, is more conveniently expressed in terms of the brightness temperature, $T_b$:
\begin{equation}
T_b(\nu, z)=\frac{I(\nu, z) c^2}{2 k_B\nu^2}.
\end{equation}

\begin{figure*}
\begin{center}
\includegraphics[width=1\textwidth]{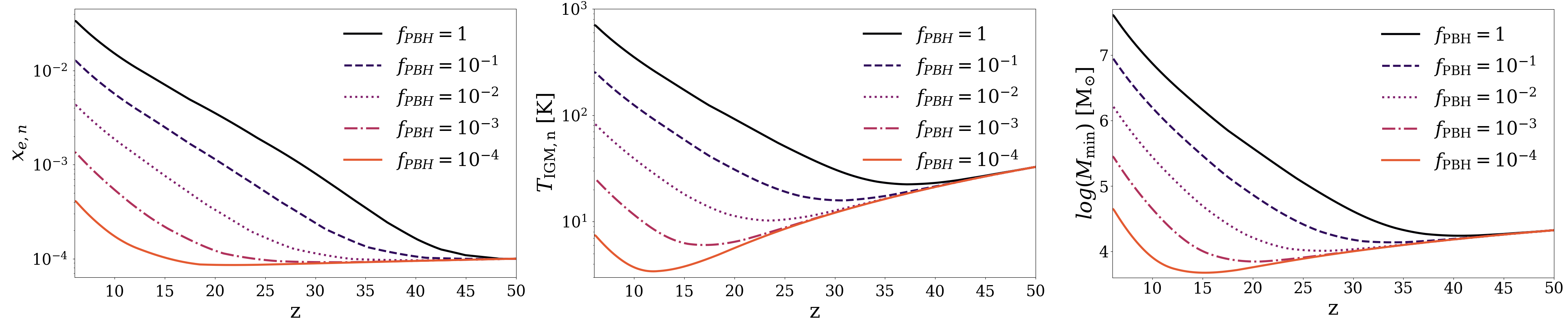}
\caption{IGM ionized fraction (left panel), Temperature (middle panel), Minimum mass (right panel) evolution with redshift inside neutral regions. Different colors represent different $f_{\rm PBH}$ values: black continuos line $f_{\rm PBH}=1$, violet dashed line $f_{\rm PBH}=10^{-1}$, purple dotted line $f_{\rm PBH}=10^{-2}$, red dashed-dotted line $f_{\rm PBH}=10^{-3}$, and orange continuous line $f_{\rm PBH}=10^{-4}$. We maintain this line style and color code thorough the rest of this paper. In all three panels a PBHs mass of $30\ M_{\odot}$ is assumed. 
}\label{IONTEMPEVO}
\end{center}
\end{figure*}
\section{IGM Heating and Ionization by PBHs}\label{heation}
Radiation emitted by accreting PBHs is injected into the IGM, thus heating and ionizing the gas. The background intensity described in the previous section can be used to self-consistently account for these large scale effects.

To compute the amount of energy injected into the IGM by the X-ray emission we follow the formalism described in \cite{Mesinger13}. The heating rate per baryon, $\epsilon_{\rm PBH}$, is a function of the X-ray background produced by PBHs, and can be expressed as:
\begin{equation}
  \epsilon_{\rm PBH}(z)=\int_{\rm \nu_{\rm min}}^{\infty}d\nu \frac{4\pi I_{\rm \nu}}{h\nu}(h\nu-E^{th})f_{\rm heat}\sigma_H(\nu),
\end{equation}
where $h\nu_{\rm min}=0.5\ \mathrm{keV}$, $I_{\rm \nu}$ is the angle-averaged specific-intensity X-ray background, $E^{th}$ is the ionization threshold, $f_{\rm heat}=0.3$ is the fraction of primary electron energy going into heat \citep{Furlanetto2010}, and $\sigma_H$ is the hydrogen ionization cross section. 

%While the main contribution to the reionization comes from stars in massive halos, the X-ray heating and ionization coming from PBHs impact mainly the neutral regions.

We then compute the impact of X-ray photons from accreting PBHs on the heating\footnote{We neglect the impact of UV ionizing radiation emitted by PBHs (see Appendix \ref{noION})} and ionization of the IGM, following \citep{Puchwein19}. The evolution of the IGM ionized fraction $x_e$ and temperature $T_{\rm IGM}$ are computed by solving the following coupled differential equations:
\begin{subequations}
\begin{equation}
  \frac{dx_e}{dz}=\frac{dt}{dz}[\Gamma - \alpha_R(T) C x_e^2 n_b],
\end{equation}
\begin{equation}
\begin{split}
  \frac{dT_{\rm IGM}(z)}{dz}=&\frac{2}{3k_B(1+x_e)}\frac{dt}{dz}\sum_i \epsilon_i\\
  &+\frac{2T_{\rm IGM}}{3n_b}-\frac{T_{\rm IGM}}{1+x_e}\frac{dx_e}{dz},
\end{split}
\end{equation}
\end{subequations}

where $\Gamma$ is the photoionization rate, $\alpha_R$ is the case A recombination coefficient , $C=2$ is the clumping factor \citep{Mesinger13}, $n_b$ is the mean baryon density, and $\epsilon_i$ is the heating rate per baryon per process $i$. 

To properly evaluate PBHs accretion in the pre-overlap phase of the cosmic reionization process, we solve the above coupled differential equations separately for the ionized and neutral regions. The assumption of volume-averaged values for the IGM temperature and ionized fraction would lead to a miss evaluation of the gas accretion onto PBHs, and subsequently of their emission. 
\subsection{Ionized regions}
We fix the IGM temperature in the ionized regions to $T_{\rm IGM, ion}\ =\ 10^4\ \mathrm{K}$; the coupled differential equations then become:
\begin{subequations}
\begin{equation}
  \frac{dx_{\rm e, ion}}{dz}=\frac{dt}{dz}[\Gamma - \alpha_R C x_{\rm e, ion}^2 n_b],
\end{equation}  
\begin{equation}
  \frac{dT_{\rm IGM, ion}(z)}{dz}= 0.
\end{equation}
\end{subequations}
In the case of ionized regions, PBHs accretion is strongly suppressed because of the increased sound speed. The redshift evolution of the ionized fraction $x_{\rm e, ion}$ traces the evolution of the volume filling factor of ionised regions, namely the fraction of volume occupied by ionized regions. We rewrite the IGM number density of PBHs dividing the ionized and neutral components as follows:
\begin{equation}
  n_{\rm IGM}(z)=n_{\rm IGM}(z)x_{\rm e, ion}(z)+n_{\rm IGM}(z)(1-x_{\rm e, ion}(z)).
\end{equation}
\subsection{Neutral regions}
\label{neutral_regions}
In the case of neutral regions, the coupled equations can be written as follows:
\begin{subequations}
\begin{equation}
  \frac{dx_{\rm e, n}}{dz}=\frac{dt}{dz}[\Gamma_{\rm PBH} - \alpha_R(T_{\rm IGM, n}) C x_{\rm e, n}^2 n_b],
\end{equation}
\begin{equation}
\begin{split}
  \frac{dT_{\rm IGM, n}(z)}{dz}=&\frac{2}{3k_B(1+x_e)}\frac{dt}{dz} \epsilon_{\rm PBH}\\
  &+\frac{2T_{\rm IGM, n}}{3n_b}-\frac{T_{\rm IGM}}{1+x_e}\frac{dx_{\rm e, n}}{dz},
\end{split}
\end{equation}
\end{subequations}
where $x_{\rm e, n}$ represents the fraction of free electrons in neutral regions, and $\Gamma_{\rm PBH}$ is the ionization rate due to PBH X-ray emission. PBHs accreting in neutral regions represent the main contributors to the cosmic backgrounds, because of their lower gas temperature that both lowers the sound speed (thus increasing the accretion rate) and allows mini-halos to virialise more easily. In neutral regions, the sole heating and ionizing contribution comes from X-ray photons, since UV photons are easily absorbed by neutral hydrogen, while X-rays can penetrate larger gas column densities before getting absorbed. 

The resulting IGM ionized fraction and temperature evolution in the neutral regions is shown\footnote{We limit our analysis at redshift $z = 6$ since, according to our model \citep{Puchwein19}, this epoch sets the end of the reionization process; thus at lower redshifts the IGM is completely ionized.} in the left and middle panel of Fig. \ref{IONTEMPEVO}, respectively. This figure clearly shows that the impact of accreting PBHs on the global thermal and ionization history of the early Universe would be particularly relevant if dark matter were completely composed of PBHs ($f_{\rm PBH}=1$): in this case, the IGM ionized fraction and temperature would increase by a factor of about 50 at $z\sim 6$; for lower values ($f_{\rm PBH}=10^{-2}$) the increment of $x_e$ and $T_{\rm IGM}$ would be limited to a factor of few.

%\begin{figure}
%\begin{center}
%\includegraphics[width=0.45\textwidth]{images/Tigm.png}
%\caption{Temperature evolution with redshift inside neutral regions. Different colors represent different $f_{\rm PBH}$ values. }\label{TEMPEVO}
%\end{center}
%\end{figure}
%\begin{figure}
%\begin{center}
%\includegraphics[width=0.45\textwidth]{images/Mmin.png}
%\caption{Minimum halo mass evolution with redshift inside neutral regions. Different colors represent different $f_{\rm PBH}$ values. }\label{Mmin}
%\end{center}
%\end{figure}
%\begin{figure}
%\begin{center}
%\includegraphics[width=0.45\textwidth]{images/Ionevo.png}
%\caption{\SG{Questa figura mi sembra sacrificabile. tra l'altro non ne parli mai }Ionization evolution with redshift inside neutral regions. Different colors represent different $f_{\rm PBH}$ values. The red dashed vertical lines are there as a reference for the redshift. Those lines represent $z\ =\ 6,\ 8$.}\label{IONEVOglobal}
%\end{center}
%\end{figure}

Finally, in the right panel of Fig. \ref{IONTEMPEVO}, we show the $\rm M_{min}$ evolution with redshift taking into account the IGM temperature evolution before reionization\footnote{The minimum halo mass after reionization is fixed to $M_{\rm min}= M_h(T_{\rm vir}=10^4\ \rm K)$.}. As a consequence of X-ray heating by PBHs, also $M_{\rm min}$ increases. In particular, $M_{\rm min}$ appears to be more strongly affected by the presence of accreting PBHs: in the case of $f_{\rm PBH}=1$ ($f_{\rm PBH}=10^{-2}$), $M_{\rm min}$ increases by factor of 1000 (10). This trend is shaped by the following combined effects: on the one hand, the relation between $M_{\rm min}$ and $T_{\rm IGM}$ is nonlinear ($M_{\rm IGM} \propto T_{\rm IGM}^{3/2}$); on the other hand, an higher value of $M_{\rm min}$ shorten the integration interval in eq. \ref{bhint}, thus reducing the resulting PBH X-ray emission from eq. \ref{intensity}.

\section{Results}\label{results}
In this Section, we constrain the fraction of dark matter in the form of PBHs by computing the contribution of PBHs to the CXB, the CRB, and the 21-cm signal.
The PBH contribution to CXB and CRB can be evaluated at different epochs. Here, we compare our cumulative model predictions at $z=0$ with the observed CXB and CRB to test to which extent the extra radiation from accreting PBHs can explain the observed excesses in the CXB and CRB. Furthermore, we constrain the possible contribution of PBHs to the CRB at $z=17$, namely the epoch at which the \HI 21 cm signal detected by the EDGES experiment shows a deep absorption feature.

\subsection{X-ray background}\label{xback}
About $\sim 70 \%$ of the CXB arises from resolved sources (mostly AGN) both in the soft and hard bands \citep{Cappelluti17}. The remaining contribution from unresolved sources can be attributed to a still unknown BH population (e.g., accreting BHs in heavily obscured conditions, \citealt{Gilli2007, Treister09}) and additional unresolved sources (e.g., extended X-ray emission from galaxy clusters \citealt{Gilli99}). To test the hypothesis that PBHs provide the origin of such unresolved background, we quantify their contribution to the X-ray background in the soft ($0.5-2\ \mathrm{keV}$) and hard ($2-10\ \mathrm{keV}$) X-ray bands by using Eq. \ref{intensity}. 

%\textbf{2.2 × 10−16}, 1.5 × 10−15, and 8.9 × 10−16 erg cm−2 s−1 in the [0.5–2], [2–10], and [0.5–10] keV energy bands, respectively.

\begin{figure}

\includegraphics[width=0.47\textwidth]{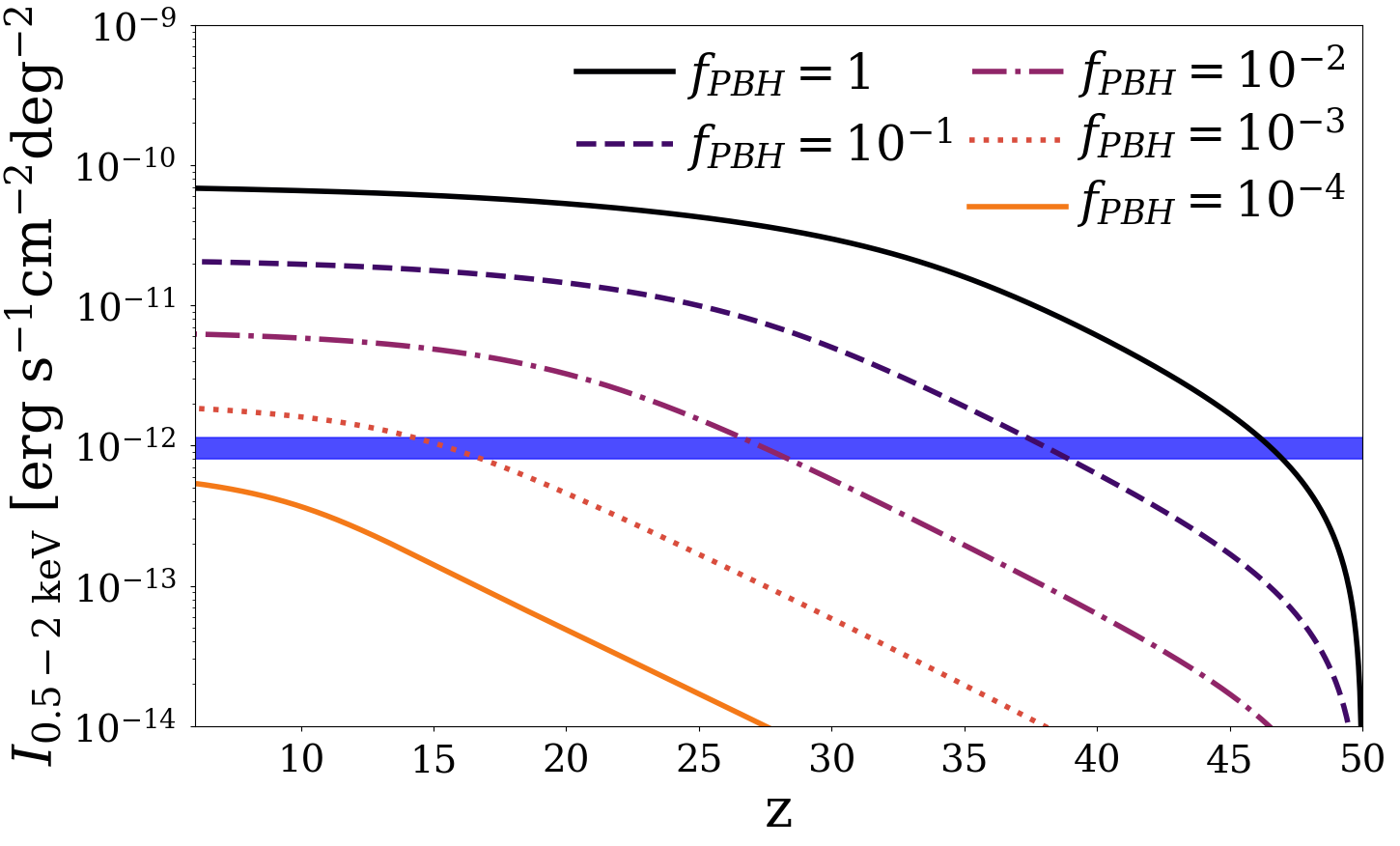}
\caption{Integrated soft X-ray background intensity $I_{0.5-2\mathrm{keV}}$ cumulative evolution, for different values of $f_{\rm PBH}$ as reported in the legend. The blue shaded regions represent the observed excess background \citep{Cappelluti21}. The background intensities are computed for $M_{\rm PBH}=30 M_{\odot}$.}\label{Xbackevo}

\end{figure}

In Fig. \ref{Xbackevo}, we compare our predictions as a function of $f_{\rm PBH}$ with observational limits reported in \citet{Cappelluti17}. The authors analyzed Chandra data \citep[COSMOS-Legacy survey,][]{Elvis09, Civano16} finding a XRB intensity at $1~ \rm keV$ to be $10.91\pm 0.16\, \rm keV\ cm^{-2} s^{-1} deg^{-2}$ and spectral index $\Gamma = 1.45 \pm 0.02$. After subtracting the X-ray detected sources they found that the unresolved CXB intensity in the soft band\footnote{In \citet{Cappelluti17} two different values are reported for the CXB, one from {\it unresolved} sources and one called {\it non-source}. The latter one refers to the observed intensity obtained when both X-ray and HST-ACS observed sources are masked. We compare our predicted value with the {\it non-source} CXB intensity.} ($I_{0.5-2 \rm keV}=2.90\times 10^{-12}\rm \ erg\ s^{-1}cm^{-2}deg^{-2}$) is smaller than in the hard band ($I_{2-10 \rm keV}=6.47\times 10^{-12}\rm \ erg\ s^{-1}cm^{-2}deg^{-2}$). We thus consider the soft band data to obtain the most stringent constraints on $f_{\rm PBH}$.
%The resulting energy spectral index is $\Gamma = 1.7$ while the observed one has a slope of $\Gamma = 1.57$. 
%The observed background intensity rapidly drops in the soft X-ray band, resulting in a stronger constraint on $f_{\rm PBH}$.

For high values of $f_{\rm PBH}$($\gtrsim 0.1$), the contribution of accreting PBHs to the soft CXB rises steeply at early epochs: for example, assuming $f_{\rm PBH}=1$, the X-ray emission from $z>40$ PBHs already overshoots the observed CXB. The integrated contribution of PBHs flattens below a critical redshift which depends on $f_{\rm PBH}$. For example, for $f_{\rm PBH}=10^{-1}$ ($f_{\rm PBH}=10^{-3}$) the PBHs cumulative emission reaches the value of 1 (10) $\rm erg~s^{-1}~cm^{-2}~deg^{-2}$ at $z\sim 20$ ($z\sim 10$) and remains almost constant afterwards. This trend is related to the redshift evolution of $M_{\rm min}$ (right panel of Fig. \ref{IONTEMPEVO}), that rises more rapidly with $f_{\rm PBH}$, as a consequence of a more efficient X-ray heating from PBHs. We find that our predictions are consistent with Chandra data only if $f_{\rm PBH} \leq 3\times 10^{-4}$. Independently from $f_{\rm PBH}$, the PBHs contribution to the XRB drops after the reionization epoch ($z<6$), since both the high gas temperature ($>10^4$~K), and the small filling factor of neutral regions make PBHs accretion extremely inefficient. 

\begin{figure}
\begin{center}
\includegraphics[width=0.47\textwidth]{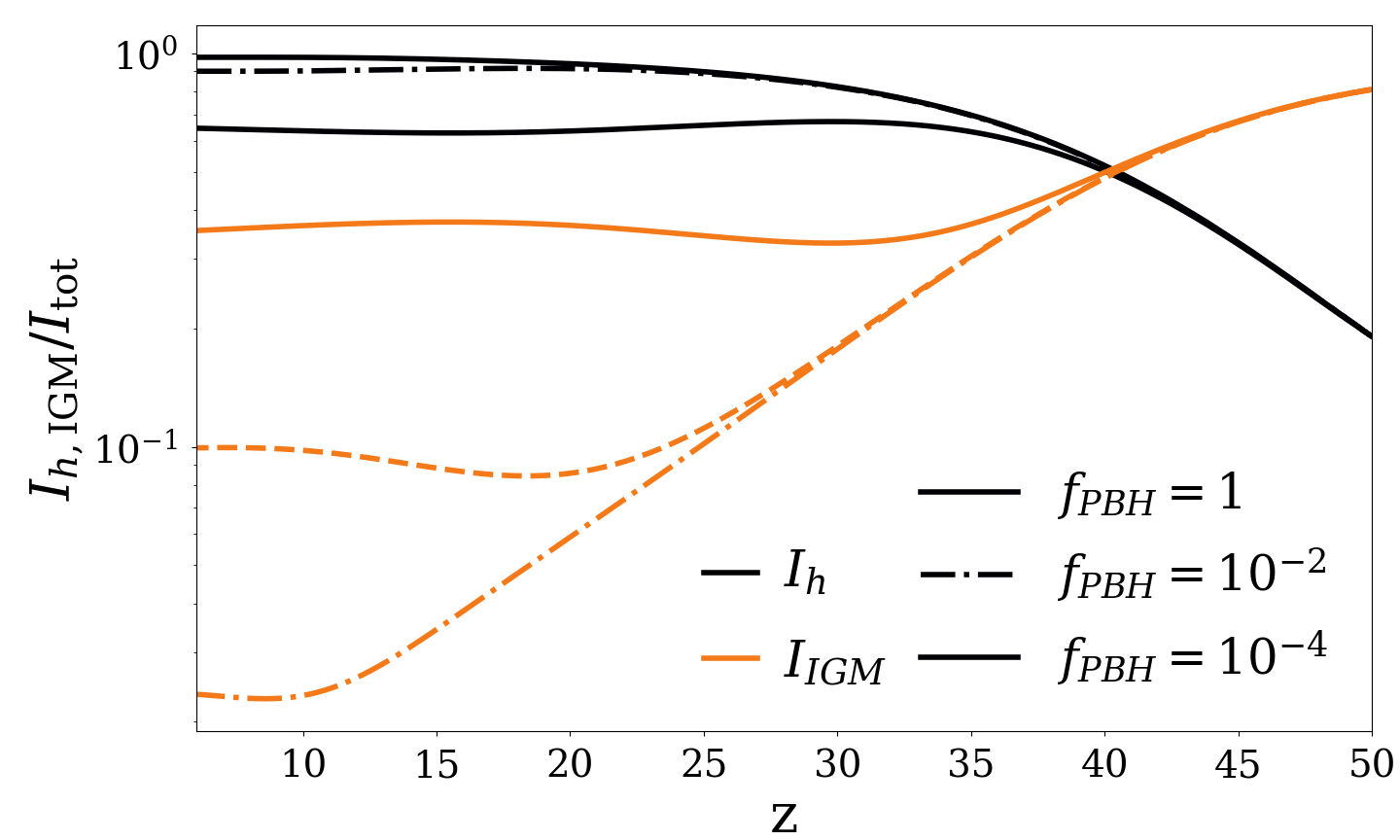}\caption{Redshift evolution of the relative contribution to the soft ($0.5-2$ keV) X-ray background from halos and IGM, $I_{h}/I_{\rm tot}$ and $I_{\rm IGM}/I_{\rm tot}$, respectively. Different line styles correspond to different $f_{\rm PBH}$, as shown in the label. Black (orange) lines correspond to the halos (IGM) contribution. The relative background intensities are computed for $M_{\rm PBH}=30 M_{\odot}$. }\label{relfluxes}
\end{center}
\end{figure}

In Fig. \ref{relfluxes} we present the redshift evolution of the ratio between the soft X-ray intensity arising from halos ($I_{X, h}$, black line) or from the IGM ($ I_{X,\rm IGM}$, orange line) and the total intensity($ I_{X,\rm tot}$). At very early epochs ($z>40$), the total signal is always dominated by IGM contribution. At $z\sim40$, the two curves cross each other, as density perturbations start to collapse into virialized structure. At later epochs, 
the total X-ray intensity is dominated by emission from PBHs in DM halos, whereas PBHs in the IGM provides only a minor contribution ($\lesssim$ 5\%). We stress that 99\% of the signal arises from mini-halos ($M_h \leq 10^6\ M_{\odot}$), whose X-ray emissivity peaks at $z\sim 6$. Independently from the value of $f_{\rm PBH}$ the total signal is dominated by the emission produced into halos, up to a maximal ratio of $I_{h}/I_{\rm IGM}\sim 50$ for $f_{\rm PBH}=10^{-4}$. The contribution from PBH accreting in the IGM increases with $f_{\rm PBH}=1$, although remaining subdominant, as a consequence of the X-ray heating feedback: as the minimum mass increases, $f_{\rm coll}$ decreases, thus raising the PBHs number density into the IGM.

In Table \ref{tabX}, we report our prediction for the background intensity in the soft and hard X-ray bands for different values of $f_{\rm PBH}$\footnote{As discussed in Sec. \ref{luminosity} the spectral index adopted for this work in the X-ray band is $\alpha = 0.7$, which is slightly steeper than the one corresponding to the background excess $\alpha = 0.57$.}. The predicted intensity appear to be stronger in the soft band for $f_{\rm PBH} \leq 10^{-3}$, and vice versa. We attribute this effect to the exponential cut-off (at $200\ \rm keV$) in the X-ray spectrum that suppresses contribution from high redshift to the hard band. In the case of a shallower intensity evolution (high values of $f_{\rm PBH}$, see Fig. \ref{Xbackevo}) a non-negligible fraction of the total radiation in the hard band is produced in the redshift interval $20\simlt z \simlt 30$ . For $f_{\rm PBH} \leq 10^{-3}$ the radiation feedback is not strong enough to suppress the signal at early epochs ($20\simlt z \simlt 30$) resulting in a steeper evolution of the CXB intensity, and consequently a lower contribution to the hard band from high redshift ($z>20$).
%At lower values, the heating kicks in at much lower redshift giving the time to mini-halos to emit on a broad redshift span, flattening the curves. For any input value of $f_{\rm PBH}$, we require $I_{\rm [0.5, 2]} \leq I_{\rm [0.5, 2], obs}$. 
%Due to the evolution of $M_{min}$ with redshift, showed in the right panel of Fig. \ref{IONTEMPEVO}, we expect also the value $M_{max}$ for which the integral converges to evolve with redshift. 

\begin{table}
\centering
\caption{X-ray background intensity in units of $10^{-12}\mathrm{erg\ s^{-1}\ cm^{-2}\ deg^{-2}}$. We report our predictions for soft ($0.5-2$ keV) and hard ($2-10$ keV) X-ray bands for the values of $f_{\rm PBH}$ considered in this work.}
\label{tabX}
\begin{tabular}{|l|c|r|}
\hline
$f_{\rm PBH}$ & $I_{0.5-2\mathrm{keV}}$ & $I_{2-10\mathrm{keV}}$ \\
\hline
\hline

$1$  & $68.2$ & $81.2$ \\

$0.1$ & $20.0$ & $22.2 $\\

$0.01$ & $6.01$ & $6.21$\\

$10^{-3}$ & $1.80$ & $1.74$ \\

$10^{-4}$ & $0.52$ & $0.47$ \\
\hline
\end{tabular}
\end{table}

\subsection{Radio background}
\begin{figure}
\begin{center}
\includegraphics[width=0.45\textwidth]{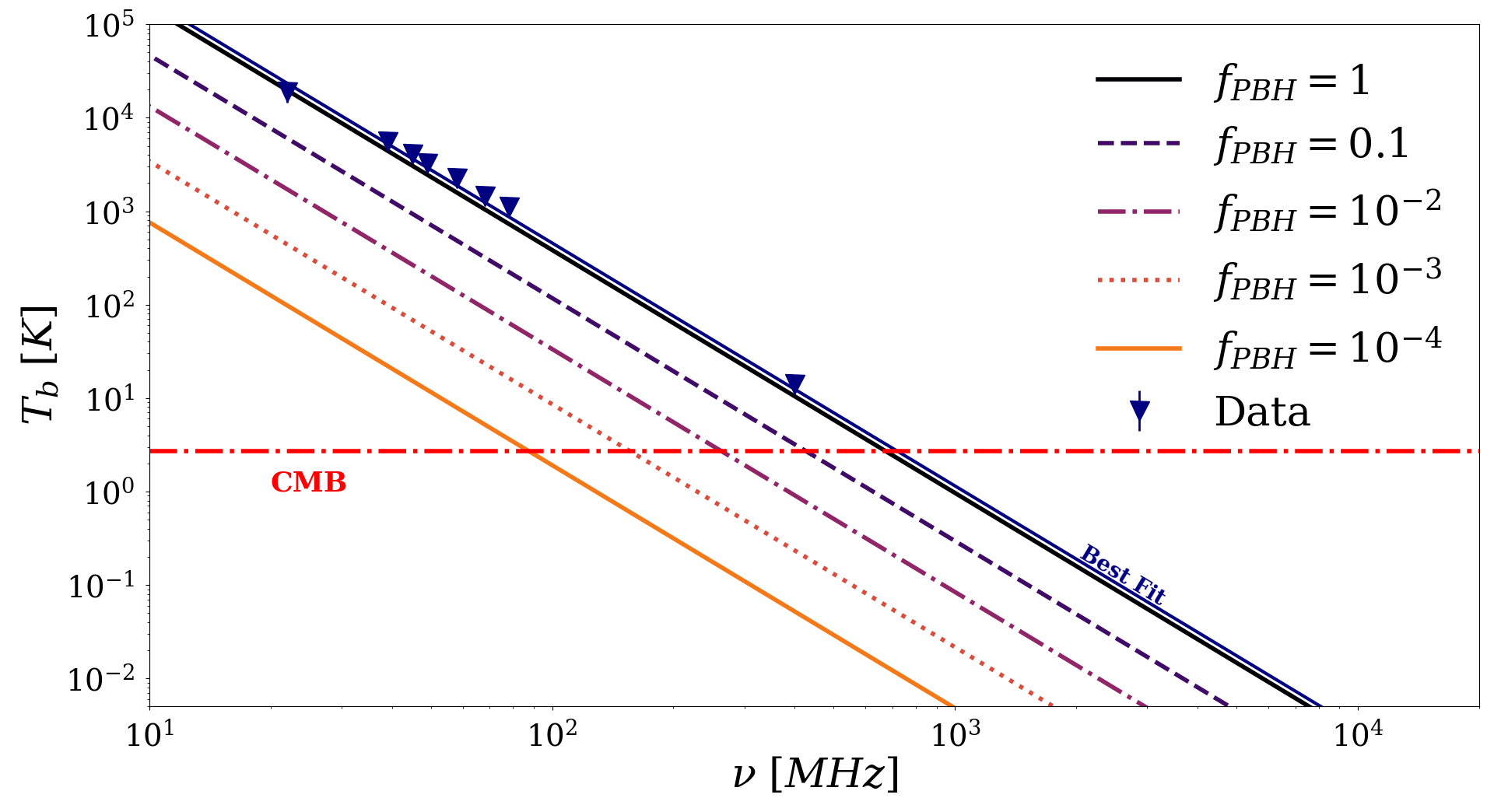}
\caption{Radio background brightness temperature at $z=0$ as a function of frequency for the values of $f_{\rm PBH}$ considered in the work. The blue triangles represent the data points \citep{ARCADE2}; the continuous blue line is the best fit for the same data. The horizontal red dotted dashed line shows the CMB temperature. The brightness temperature is computed at $1.4\ \rm GHz$ for PBHs of $30\ M_{\odot}$.}\label{myarcade}
\end{center}
\end{figure}

The observed background radio-excess is well fitted with a power-law spectrum between 22 MHz and 10 GHz \citep{ARCADE2}:
\begin{equation}
T_b=(24\pm 2.1\rm K) \times \left(\frac{\nu}{310\ \rm MHz}\right)^{-2.599\pm0.036}.
\end{equation}
In our model, the X-ray emission in the accretion process is associated to jet production and thus to radio emission (see Sec. \ref{luminosity}). We thus compute the contribution of PBHs to the radio background at 1.4 GHz, at redshift $z = 0$, and compare our results with ARCADE2 data ($480~\rm mK$). In Fig. \ref{myarcade}, we show the brightness temperature of the radio background, as predicted by our calculations, at varying $f_{\rm PBH}$. The amplitude of the radio background increases sub-linearly with $f_{\rm PBH}$ (see also Tab. \ref{tabR}), while the shape assumed for synchrotron emission fairly reproduces observations, and does not depend on $f_{\rm PBH}$.

%X-ray are overproduced with respect to radio, due to the nature of the fundamental plane relation. This allows us to fix $f_{\rm PBH}$ and cut off a region of the parameter space. 
In the case $f_{\rm PBH} = 1$, the PBHs contribution to the CRB would represent $92 \%$ of the observed excess, thus explaining most of the unexplained signal. However, if we take into account CXB constraints, PBHs cannot explain the radio excess, since they would contribute only 1\% to the CRB. 

%\SG{Da rivedere Fig. \ref{XradioFDM} shows how the X-ray and radio background produced by PBHs are subject to those constraints. The amount of radiation produced in the radio band is fixed by the lower limit of X-rays which does not exceed the observed background value. In the plot we show the fraction of allowed background, $f_{\rm back}$, versus $f_{\rm PBH}$. From the plot is clear that for the signal produced by PBHs cannot reproduce the entirety of the ARCADE2 signal even for values of $f_{\rm PBH}$ higher than the ones allowed by the X-rays constraint. As expected, due to the nature of the fundamental plane relation X-rays are strongly overproduced with respect to radio. This results in a disparity that brings to the impossibility to fully match the radio observation made with ARCADE2. It appears impossible to recover $f_{\rm back}=1$ for the radio background. This trend is also expected due to the lack of absorption in the environment close by PBHs. Mini halos, in which the vast majority of the signal is generated, are not dense enough to absorb a substantial fraction of the high energy radiation.}  
\subsection{21 cm signal}
The radio background due to accreting PBHs may affect the amplitude of the 21-cm global signal. The absorption depth of the signal can be computed as follows \citep{Mesinger13}:
\begin{equation}
\begin{split}
  \delta T_b = & 27 x_{\rm HI} (1+\delta) \biggl(\frac{\Omega_b h^2}{0.023} \biggl)\biggl(\frac{0.15}{\Omega_m h^2}\frac{1+z}{10}\biggl)^{1/2} \\
  & \frac{\partial_{r}v_r}{H(z)1+z}\biggl(1-\frac{T_r}{T_s}\biggl),
\end{split}
\end{equation}
where $x_{\rm HI}=1-x_e$ is the neutral hydrogen fraction, $\partial_{r}v_r$ is the comoving gradient of the line of sight component of the comoving velocity, $T_r$ is the radio background temperature, $T_s$ is the spin temperature. \citet{fialkov19} pointed out that, given a radio background of amplitude $\rm A_r$ in addition to the CMB, the radio background temperature can be written as follows:
\begin{equation}
  T_r=T_{\rm CMB}(1+z)\biggl[1+A_r\biggl( \frac{\nu_{\rm obs}}{78\ \rm MHz}\biggl)^{-2.6} \biggl],
\end{equation}
where $\nu_{\rm obs}$ is the observed frequency by the EDGES instrument at $78\ \rm MHz$, which corresponds to $z = 17$. Thus, the presence of a radio background in addition to the CMB, may push the absorption signal to larger depths, possibly explaining the results obtained with the EDGES experiment \citep{EDGES}. In particular, to recover an absorption signal consistent with EDGES, an extra radio background with an amplitude $A_r > 1.9$ is required. 

In Tab. \ref{tabR}, we report our predictions on the brightness temperature of the 21 cm at $z = 17$, as predicted by our calculations, varying $f_{\rm PBH}$: for $f_{\rm PBH}=1$ ($f_{\rm PBH}=10^{-4}$), we obtain $A_r=10$ ($0.01$); in order to reproduce the entire EDGES absorption feature, $f_{\rm PBH} \gtrsim 4\times 10^{-2}$ is required. However this value would exceed the constraints derived from the CXB. The maximum value allowed by the CXB provides $T_R=0.59\ \rm K$: this radio background temperature can explain only $1\%$ of the entire EDGES signal. 

%It seems impossible to recover the entire absorption signal while respecting the X-ray constraints discussed in the previous section. The only allowed value $f_{\rm PBH}=10^{-4}$ with the associated temperature $ \rm T_r=16\ \rm K$ represent the $17\%$ of the minimum value $\rm T_r = 1.9\ T_{CMB}$. The different cases for $\rm f_{\rm PBH}$ represent respectively $\rm A_r=0.3,\ 1.7,\ 6.9,\ 23,\ 64$. 

%We thus conclude that, for the assumed mass distribution, PBHs are unable to reproduce the background intensity needed to justify the EDGES feature if we impose the constraint derived in sec \ref{myconst} for the X-rays. In the case of $\rm f_{\rm PBH} \gtrsim 10^{-3}$ the extra radiation would reproduce a signal deep enough to match the EDGES feature. 
We point out that our result can be modified by further modeling the physics of the 21-cm signal (e.g. the impact of PBHs radiation on $T_s$ via Lyman-$\alpha$ coupling). Even if our results only provide a first order of magnitude estimate, we do not expect the PBHs contribution to the 21-cm signal to be relevant even with further improvement of the model. 

\begin{table}
\centering
\caption{Predictions for the radio background brightness temperature (in units of [K]), both at $z = 0$, and $z = 17$ for the values of $f_{\rm PBH}$ considered in this work.}
\label{tabR}
\begin{tabular}{|l|c|c|r|}
\hline
$f_{\rm PBH}$ & $T_b(z=0)$/K & $T_b(z=17)$/K & $A_r$ \\
\hline
\hline
$1$  & $0.402$ & $579.67$ & $11.80$\\
$0.1$ & $0.123$ & $174.42$ & $3.55$ \\
$0.01$ & $0.035$ & $42.81$ & $0.87$ \\
$10^{-3}$ & $0.009$ & $7.59$ & $0.15$\\
$10^{-4}$ & $0.002$ & $1.10$ & $0.02$ \\
\hline
\end{tabular}
\end{table}
\section{PBHs as dark matter: constraints}\label{PBHconst}
\begin{figure*}
\begin{center}
\includegraphics[width=1\textwidth]{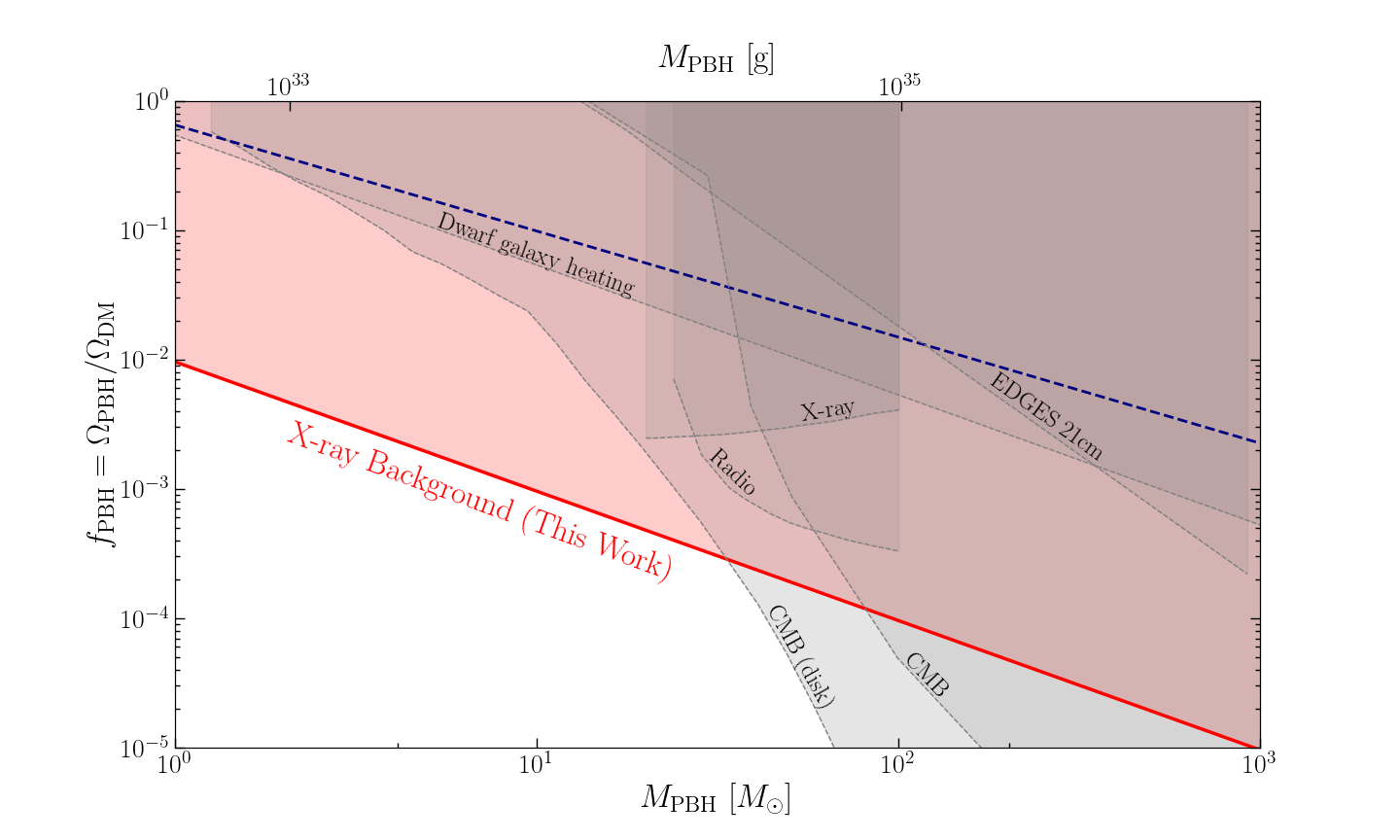}
\caption{Constraints on $f_{\rm PBH}$ derived taking into account PBHs accretion. The red solid line represents the constraint obtained in this work from the cosmic X-ray background. The blue dashed line defines the minimum $f_{\rm PBH}$ value to recover the EDGES signal depth. Dashed gray lines correspond to existing constraints: (a) dwarf galaxy heating: derived by evaluating the impact of gas heating due to PBHs accretion in dwarf galaxies \citep{Lu20}; (b) X-ray and radio: obtained through the comparison of the number of X-ray and radio emitters in the Milky way center and the predicted number by a PBHs accreting model from \citet{Manshanden19}; (c) CMB: derived by evaluating the impact of radiation emitted by PBHs on CMB anisotropies assuming spherical or disk accretion \citep{Serpico20}; (d) EDGES 21cm: computed by considering the impact of heating feedback from PBHs accretion on the IGM, and comparing the results with the predicted value \citep{Hektor18}. Figure created by modifying the publicly available Python code PBHbounds \citep{Kavanagh19}.}\label{constraint}
\end{center}
\end{figure*}

Fig. \ref{constraint} summarises the main results of this work and compares them with previous ones in the literature.
Dashed lines in the plot show existing constraints from the accretion process. The line labelled "dwarf galaxy heating" refers to the results obtained in \citet{Lu20} through the analysis of gas heating due to PBHs accretion. \citet{Manshanden19} modeled the X-ray and radio emission from accreting PBHs in the Milky Way. Source counting in those bands only allows for a small fraction ($f_{\rm PBH} \leq 10^{-3}$) in the $10 \leq M_{\odot}\leq 100$ mass range. 

The EDGES-21cm marks the constraints by \citet{Hektor18}, who evaluated the impact of heating and ionization from PBHs accreting in the IGM only\footnote{We underline that in our work we instead take into account PBH accretion in halos as well.}, by imposing that $T_{\rm IGM}(z=17)\leq 8\ \rm K$. 
 
The most stringent constraint in the considered mass range ($1 M_{\odot} \leq M_{\rm PBH} \leq 10^3 M_{\odot}$) has been set by \citet{Serpico20}, who evaluated the impact of PBHs energy production on CMB anisotropies. In their work they considered two different models for PBHs accretion, assuming either spherical or disk accretion, finding two different upper limits on $f_{\rm PBH}$. In the case in which PBHs are not the sole constituent of DM, $f_{\rm PBH} < 1$, the authors modeled PBH-DM interaction, and assumed a dark matter halo to form around each PBH. This effect was already discussed in \citet{ricotti08}, and it results in a time-increasing halo mass surrounding the PBH, $M_{h}(z)=(3000/1+z) M_{\rm PBH}$. The result is a boost factor of PBH accretion rate of $\sim 10^2$ at ($z=10$). We do not account for PBH-DM interaction in the case $f_{\rm PBH} < 1$, and note that accounting for it would strengthen our results, boosting PBHs accretion and consequently the CXB and CRB contributions. Even in the absence of PBH-DM interaction we can set the strongest constraint at $M_{\rm PBH}=30 M_{\odot}.$

We show in Fig. \ref{constraint} the upper bound on $f_{\rm PBH}$ as a function of PBHs mass, and the new, strong constraint $f_{\rm PBH}\leq 3\times 10^{-4}(30 M_{\odot}/M_{\rm PBH})$ from the X-ray background. The dependence $M_{\rm PBH}^{-1}$ follows from the halo X-ray luminosity relation, $L_X \propto M_{\rm PBH}^2/M_{\rm PBH}\propto M_{\rm PBH}$, where the factor $M_{\rm PBH}^2$ ($1/M_{\rm PBH}$) arises from black hole accretion (number density of PBHs in the halo). In Fig. \ref{constraint} we also show the lower limit required to recover the EDGES depth as a blue dashed line. We note that the entire region is excluded by the X-ray background constraint.
We stress that we set the strongest constraint on $f_{\rm PBH}$ by comparing the predicted value of the soft CXB with the observed one. PBHs are thus unable to contribute significantly to the CRB at any epochs, and consequently their emission is unable to affect the global 21-cm signal.

We briefly comment on the implication of the new constraint on extended PBHs mass distributions (e.g. \citealt{Bellido19, Jedamzik21}). Several multi-peaked mass functions present their main peak at $\sim 1\ M_{\odot}$. Constraining the low-end of the solar mass window implies that we limit the height of the main peak at $f_{\rm PBH} \sim 10^{-2}$, value which is typically overshoot. This would imply a reconsideration of some properties (e.g height and location of the main peak) for those mass distributions. We stress that to properly evaluate the impact of our analysis on extended mass distributions the approximation of a monochromatic PBHs spectrum has to be relaxed.

\section{Summary}\label{conclusion}
In this work, we presented a new semi-analytical model to compute the accretion rate and the relative luminosity of PBHs in dark matter halos and in the IGM, and explored the possibility that radiation emitted by accreting PBHs could contribute to the X-ray (CXB) and radio (CRB) backgrounds. Our main results are the following:
\begin{enumerate}
\item By comparing the cumulative X-ray emission from PBHs predicted by our model with the observed CXB, we set the strongest existing constraint on $ f_{\rm PBH} \leq 3\times 10^{-4}\ (30/M_{\rm PBH})$ in the mass range $1-1000\, M_\odot$.
\item We compare the PBH radio emission predicted by our model with the measured 21cm line feature and CRB. In the first case the extra radio background is able to recover only a fraction $f = 1 \%$ of the absorption depth claimed by \citet{Bowman18}. 
Similarly, the allowed brightness temperature $T_b(\rm 1.4~GHz)\sim 0.004~K$ represents only a fraction $f \sim 1\% $ of the radio excess observed by ARCADE2.
\item Depending on the minimum halo mass considered, we show that the contribution from PBHs accreting into DM halos dominates over those accreting from the IGM ($60\%<I_{h}/I_{\rm tot}<99\%$).
\item Most of the CXB/CRB emission is produced by PBHs in DM mini-halos ($M_h \leq 10^6\ M_{\odot}$) at early epochs ($z>6$).
\item We also evaluated the impact of PBHs radiation on the thermal state of the IGM. We find that the X-ray heating impacts the IGM temperature in neutral regions and consequently the minimum halo mass. This effect introduces a nonlinear dependence of the background intensity on $f_{\rm PBH}$. 
\end{enumerate}

As a final caveat, we stress that we have assumed a monochromatic PBH mass function; in future work this assumption can be relaxed. An additional assumption concerns the use of a standard $\Lambda \rm CDM$ halo mass function. For consistency, one should include the modifications induced by the presence of a fraction $f_{\rm PBH}$ of dark matter in the form of PBHs. We do expect this inclusion to boost the number of mini-halos. As these are the major contributors to the CXB and CRB, we predict that the upper limits on $f_{\rm PBH}$ could be pushed to even smaller values.

\section*{Data Availability}
Data generated in this research will be shared on reasonable request to the corresponding author.

\section*{ACKNOWLEDGEMENTS}
We thank A. Mesinger, A. Pallottini, Y. Qin for useful discussions.

\appendix\label{column}
\section{Halo concentration}\label{concentration}
Halo concentration describes the central density of an halo in terms of its mean density. It can be expressed as a ratio of radii, mean velocities or densities. It plays a key role in shaping the dark matter density distribution, and consequently the baryon density distribution. The relation adopted in this work (Eq. \ref{maccio}) allows us to compute the concentration parameter at $z=0$. We modeled the redshift evolution as $c\propto(1+z)^{-1}$, as suggested by previous works \citep{barkana02,duffy08,leoT}. 

\begin{figure}
\includegraphics[width=0.48\textwidth]{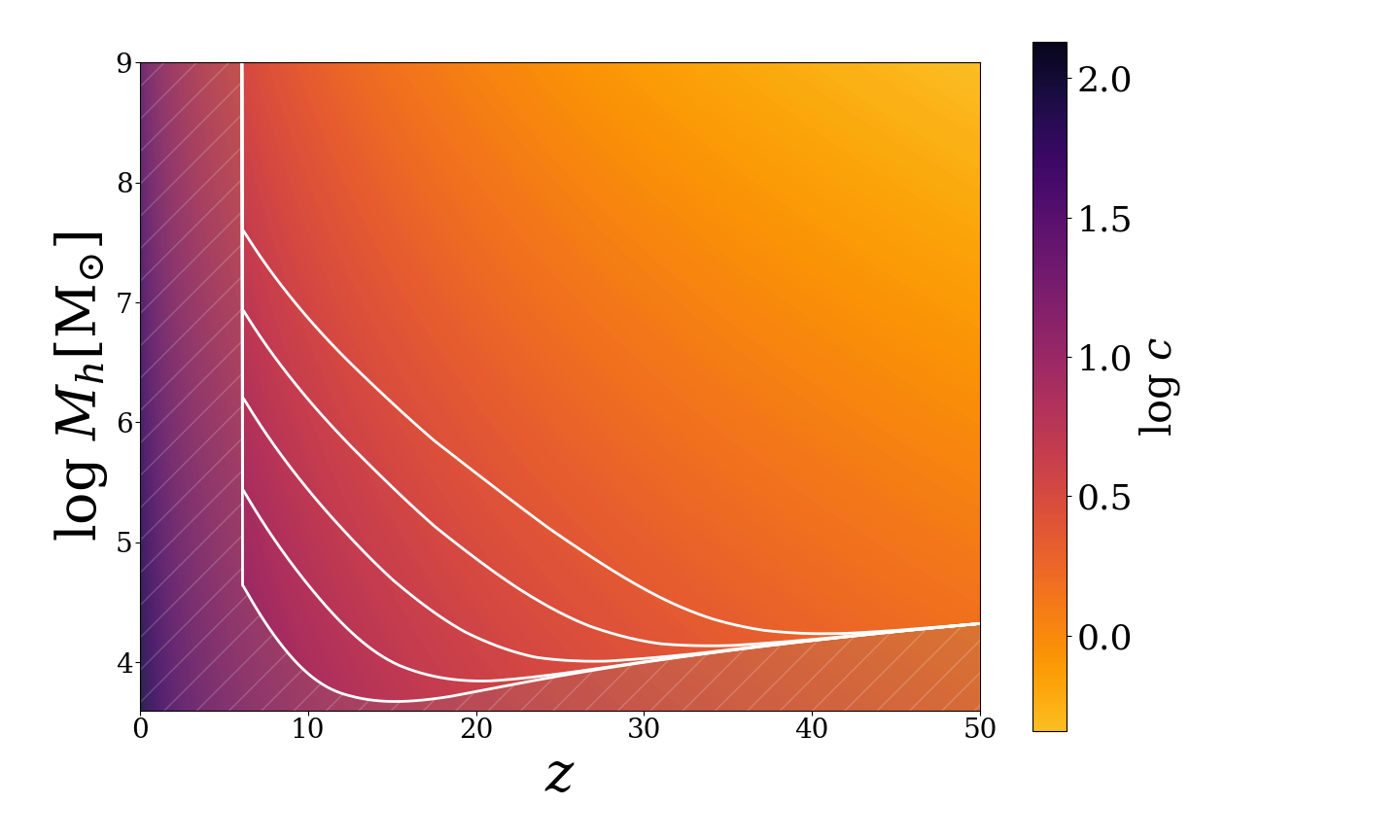}
\caption{Evolution of the concentration parameter $log\ c$ with mass and redshift. The white continuous lines represent different minimum halo masses as a function of $f_{\rm PBH}=1, 0.1, 10^{-2}, 10^{-3}$.}\label{logc}
\end{figure}

The mass-redshift evolution of the concentration parameter is shown in fig \ref{logc}. The figure is the result of an extrapolation for low mass and high redshift of eq. \ref{maccio}. 
The concentration-mass-redshift relation for dark matter halos is still an open problem. The general approach to derive this relation is to observe the behaviour of halo concentration in simulations, for a fixed range of mass and redshift \citep{maccio07,Klypin16,Ludlow16}. It has been found that the concentration decreases with both redshift and mass, $c\propto M^{-\alpha}(1+z)^{-\beta}$ \citep{Bullock17,Bullock01}. Besides this general trend, several controversies still exist. 
In \citet{prada12} was pointed out that a more general evolution can be obtained relating the concentration parameter to the r.m.s. fluctuation of the density field, $c(\sigma(M, z))$. This also implied an increasing trend with the halo mass/redshift supported by different works \citep{maccio14,Klypin11}. The apparent tension between those works and previous results was relaxed arguing that the effect appears due to the inclusion of non-relaxed halos in the sample \citep{Ludlow12}. Part of the disagreement in the high-mass end is caused by the comparison of concentration based on radial ratios and concentration based on velocity ratios. 

Different models also disagree on the concentration parameters of light-halos ($M < 10^8\ M_{\rm \odot}$). Deriving halo concentration parameter from simulations suggests a power-law dependence both on redshift and mass. The resulting relations are valid in a fixed range of mass and redshift, extrapolating beyond the validity range often results in extremely high concentration values ($c > 100$) for very small halos ($M \sim 10^{-5} M_{\rm \odot}$). Compact mini-halos can results in two effects: amplification of the annihilation rate, and overestimation of the boost factor due to substructure. The former has been observed in works relative to dark-matter annihilation resulting in $\gamma$-ray emission (e.g., \citealt{Ando19}). The latter one has been addressed in \citet{prada14.2}, clarifying that a more conservative evolution of the concentration parameter \citep[e.g.][]{prada12}) can reduce this effect.
We avoid this criticality since the region of the parameter space ($M, z$) in which the effect is dominant is always excluded by the evolution of the minimum halo mass.

\section{Ionizing photons from PBHs}\label{noION}
\begin{figure}
\includegraphics[width=0.5\textwidth]{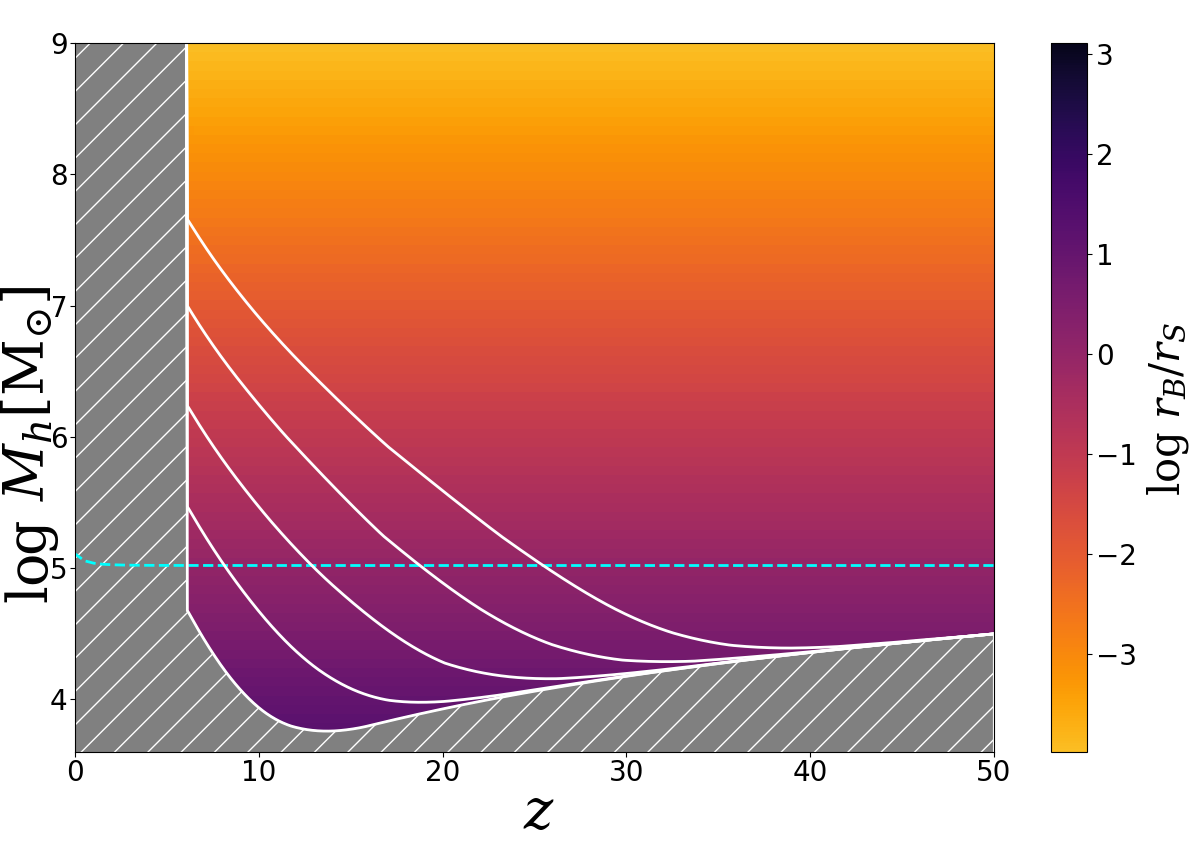}
\caption{Evolution of the ratio between the Bondi radius, $r_B$, and the Str\"omgren radius, $r_S$, as a function of halo mass and redshift. The computation is performed at a halo radius $r = 0.01\ r_{vir}$, and for a UV bolometric correction $f_{UV}=0.01$. The dashed cyan line represents the case $r_B/r_S = 1$. The white continuous lines represent different minimum halo masses as a function of $f_{\rm PBH}=1, 0.1, 10^{-2}, 10^{-3}$.}\label{bondistromg}
\end{figure}

In this paper we neglect the effects of UV ionizing photons emitted by PBHs. To motivate this choice, we start from Fig. \ref{bondistromg}, where we show the ratio between the Bondi radius and the Str\"omgren radius as a function of halo mass and redshift. From this figure, it is possible to determine whether UV photons are trapped inside the Bondi radius. We find that for very high accretion rates, $\dot{M} \gtrsim 10^{-1}\dot{M}_{E}$, UV photons cannot travel beyond the Bondi radius. This condition is verified in low-mass halos. Due to their large number density, they are the largest contributors to the cosmic backgrounds. This motivates our choice to neglect UV ionizing radiation in the model. A similar argument was presented in \citep{Hasinger2020}. Even in the case of UV photons escaping the Bondi radius, the density of the surrounding medium is high enough to result in a fast hydrogen recombination. The condition $t_H/t_{r} > 1 $ has to be verified for photons to escape the proximity of PBHs, where $t_H$ is the Hubble time, and $t_r$ is the recombination time. 

%\textbf{On the other hand, including the impact of UV ionizing photons in the model would allow us to explore different accretion scenarios (e.g. \citealt{PR1, PR3}), in which the ionization feedback is crucial. }
%\addcontentsline{toc}{chapter}{References}
%\bibliographystyle{file_bibliography/apj}

\bibliographystyle{mnras}
\bibliography{Bibliography/biblio}
\bsp
\label{lastpage}

\end{document}